\documentclass[final,5p,times,twocolumn]{elsarticle}

\usepackage{amssymb}
\usepackage{amsmath}

\usepackage{float}
\usepackage{amsfonts}
\usepackage{algorithmic}
\usepackage{graphicx}
\usepackage{subfigure}
\usepackage{textcomp}
\usepackage{xcolor}
\usepackage{booktabs}
\usepackage{colortbl}
\usepackage{threeparttable}
\usepackage{color}
\usepackage{multirow}
\usepackage{url}
\usepackage{enumitem}
\usepackage[linesnumbered,vlined,ruled]{algorithm2e}
\usepackage{graphicx}
\usepackage{subfigure}

\journal{Nuclear Physics B}

\begin{document}

\begin{frontmatter}



\title{Hybrid Matrix Factorization Based Graph Contrastive Learning for Recommendation System} 


\author{Hao Chen}\ead{202200358067@ytu.edu.cn}
\author{Wenming Ma\corref{cor1}}\ead{mwm@ytu.edu.cn}
\author{Zihao Chu}\ead{czh1126940012@s.ytu.edu.cn}
\author{Mingqi Li}\ead{lmq135831@s.ytu.edu.cn}
\affiliation{
    addressline={School of Computer and Control Engineering}, 
    organization={Yantai University},
    city={YanTai},
    postcode={264005}, 
    country={China}}
\cortext[cor1]{Corresponding author}

\begin{abstract}
In recent years, methods that combine contrastive learning with graph neural networks have emerged to address the challenges of recommendation systems, demonstrating powerful performance and playing a significant role in this domain. Contrastive learning primarily tackles the issue of data sparsity by employing data augmentation strategies, effectively alleviating this problem and showing promising results. Although existing research has achieved favorable outcomes, most current graph contrastive learning methods are based on two types of data augmentation strategies: the first involves perturbing the graph structure, such as by randomly adding or removing edges; and the second applies clustering techniques. We believe that the interactive information obtained through these two strategies does not fully capture the user-item interactions. In this paper, we propose a novel method called HMFGCL (Hybrid Matrix Factorization Based Graph Contrastive Learning), which integrates two distinct matrix factorization techniques—low-rank matrix factorization (MF) and singular value decomposition (SVD)—to complementarily acquire global collaborative information, thereby constructing enhanced views. Experimental results on multiple public datasets demonstrate that our model outperforms existing baselines, particularly on small-scale datasets. 
\end{abstract}




\end{frontmatter}



\section{Introduction}\label{Indro}

Information overload is a significant issue arising from the rapid development of the Internet. Users are confronted with a massive amount of information and product choices when utilizing the Internet, making it challenging to identify and resonate with content that aligns with their interests and preferences. Therefore, a mechanism is needed to assist users in filtering and recommending content that they are interested in from the vast sea of information. This need has given rise to recommendation systems\cite{ricci2010introduction}. Overall, recommendation systems have evolved from early collaborative filtering (CF) to content-based recommendations, and then to personalized and diversified recommendations. The continuous advancements in algorithms and technologies have enabled recommendation systems to meet users' personalized needs as much as possible, enhancing the rationality, accuracy, and user experience of recommendations.

In graph-based recommendation systems \cite{wang2019neural}, graph neural networks (GNN) play a critical role. The multiple message-passing layers in GNNs enable the aggregation of neighborhood information, thereby integrating contextual information and generating user and item embeddings that contain high-order interaction information between users and items \cite{chen2020revisiting}. These obtained embeddings are then used to iteratively refine the user-item interaction graph \cite{zhang2019star,liu2021interest}. In supervised learning, the data sparsity \cite{lin2021task} arises due to the lack of sufficient high-quality labeled data for training, which results in inadequate user interaction information and decreased recommendation performance. Therefore, the advantages demonstrated by contrastive learning once again provide an effective solution to this problem \cite{he2019cascade}.

Recent research has once again highlighted the advantages of contrastive learning, which has played a significant role in the field of graph machine learning due to its ability to acquire sufficient information without the need for labels or category information. Unlike traditional supervised learning, the principle of contrastive learning (CL) is to bring similar samples as close as possible and push dissimilar samples far apart, thereby maximizing similarity and enlarging dissimilarity \cite{xie2022contrastive}. It is this characteristic that differentiates CL in handling data sparsity issues, demonstrating better performance compared to other methods. Recently, CL has achieved notable success across multiple domains.

The core of contrastive learning lies in similarity and dissimilarity, whereby samples from the same class are mapped to the same feature space, and samples from different classes are mapped to distinct feature spaces. This ensures that samples within the same mapped space are close to each other, while samples in different mapped spaces are far apart \cite{wu2021self}. The most typical way to apply contrastive learning (CL) in recommendation systems is to augment the user-item bipartite graph using certain methods, such as randomly dropping a certain number of nodes or edges and maximizing the similarity of consistent representations.

Since the core issue of contrastive learning (CL) revolves around how the view generator produces and identifies contrastive samples, designing an effective view generator becomes particularly important when applying CL to enhance graph-based recommendation systems. The most common approach is to use heuristic generators for data augmentation \cite{xia2022hypergraph}. For example, LightGCL proposes collecting user-item interaction information via Singular Value Decomposition (SVD) to generate perturbed graphs \cite{cai2023lightgcl}, while SGL advocates random graph augmentation by randomly placing nodes or edges \cite{wu2021self}. NCL suggests clustering neighboring nodes to capture user-item information, ensuring that the representations of adjacent nodes in the graph are as similar as possible \cite{lin2022improving}. On the other hand, SimGCL employs noise injection to disturb the embedding representations \cite{yu2022graph}. While these methods have achieved some success, they also expose certain issues. Firstly, random augmentation techniques, though simple and efficient for generating contrastive views, suffer from problems of information loss or interference from irrelevant information. Secondly, such view generators have poor generalization, particularly for individual users, making the already scarce information even more unusable.

Given the limitations and challenges previously outlined, and inspired by LightGCL, we propose our method, HMFGCL. In our approach, we employ two techniques for graph augmentation: Matrix Factorization (MF) and Singular Value Decomposition (SVD), combining them to generate perturbed views for contrastive learning. This allows us not only to capture the fundamental user-item interaction information but also to incorporate global contextual information. Additionally, we enhance the quality of the embeddings \cite{yu2023xsimgcl} by introducing mixed noise during the propagation process of the graph neural network. As a result, our method can account for the dependencies among users.

\begin{itemize}
    \item We propose a novel recommendation method, HMFGCL, which leverages a hybrid approach of two distinct methods for guided graph enhancement to capture user-item interaction information in a complementary manner, incorporating global contextual information into the contrastive learning representation.
    \item To enhance the representation quality during the information propagation process of graph neural networks, we introduce mixed noise into the propagation procedure of the graph neural networks.
    \item Extensive experimental results on multiple public datasets demonstrate that our approach outperforms baseline methods. Subsequent experimental analysis confirms the rationality and robustness of HMFGCL.
\end{itemize}

\section{Related works}

In recent years, the use of graph neural networks (GNNs) for solving recommendation problems has seen a proliferation of research, achieving considerable success and playing an important role in various application scenarios. The key aspect of GNNs lies in information aggregation and propagation, where neighbor information is aggregated to model user preferences, and the graph is reconstructed to pass the information to the next layer for updating other nodes. Among the most representative models is the Graph Convolutional Network (GCN), which employs multiple convolutional layers to capture contextual information and thereby aggregate information from neighboring nodes. Various methods extended from GCN have achieved remarkable results in multiple fields. For instance, Zhao et al. \cite{zhao2019t} successfully predicted traffic flow by integrating Graph Convolutional Networks (GCN) with Gated Recurrent Units (GRU). Yang et al. \cite{yang2024predicting} proposed a traffic flow prediction approach that combines GCN with time series models to effectively model complex traffic data. Wang et al. \cite{wang2022towards} delved into the impact of uniformity and alignment in representation embeddings on the performance of recommendation systems using a straightforward GCN. Yu et al. \cite{yu2021self} introduced MHCN, a method for social recommendation using multi-channel hypergraph convolutional networks in a self-supervised manner. He et al. \cite{he2024graph} proposed a novel dual-tower model, IHDT, for modeling heterogeneous graphs. Meng et al. \cite{meng2024cross} analyzed the performance differences before and after the fusion of four different techniques. 

Contrastive learning has seen a resurgence in recent years, achieving remarkable results. It operates in a self-supervised manner, demonstrating advantages in addressing the issue of label sparsity. The earliest recommendation model utilizing contrastive learning is S³-Rec \cite{zhou2020s3}, which operates in a sequential recommendation scenario. It employs data augmentation through random masking and accomplishes recommendation tasks by maximizing consistency and dissimilarity. You et al. \cite{you2020graph} introduced the GraphCL framework, which incorporates four distinct graph augmentation methods. By employing various configurations, the framework achieved excellent generalization and performance. Wang et al. \cite{wang2022contrastive} proposed a general framework called CLSA, which demonstrates that strong augmentations outperform weak augmentations through a contrastive analysis between weakly augmented and strongly augmented data. Zheng et al. \cite{zheng2021weakly} introduced a weakly supervised contrastive learning framework, WCL, to address the issue of class conflict.

In recommendation systems, numerous methods have also achieved commendable results. The model proposed by NCL \cite{lin2022improving} combines MLP and GMF submodels, leveraging two different methods to obtain information. SGL \cite{wu2021self} introduced a graph enhancement method involving node dropout, but it suffers from the issue of information loss. SimGCL \cite{yu2022graph} innovatively proposed a feature perturbation method for data augmentation, adding small-scale noise to feature embeddings, which has yielded positive results. The GNNCL method proposed by Chen et al. \cite{chen2024gnncl} integrates noise and clustering into the framework of Graph Neural Networks (GNNs), achieving notable results. As a novel approach, LightGCL \cite{cai2023lightgcl} introduced SVD-guided graph enhancement, moving away from the traditional three-view paradigm and achieving success.

\section{Problem definition and preliminaries }

In this section, we present the fundamental problem definition of recommendation methods based on graph contrastive learning.

\subsection{Problem definition}
Our objective is to utilize the user-item graph $G$, the user-item adjacency matrix $A$, and two low-rank approximation matrices decomposed using different methods to train the function $F$ for graph enhancement. The low-rank approximation matrix generated by MF is denoted as $\tilde{A}$, and that by SVD is denoted as $\hat{A}$.
\begin{flalign}
    &&
    \hat{y}_{u,v}=F(u,v\mid G,A,\tilde{A},\hat{A},\Pi )
    &&
\end{flalign}
here, $u$ represents a user, $v$ represents an item, and $\hat{y}_{u,v}$ denotes the interaction probability between user $u$ and item $v$. Within the function $F$, $\Pi$ denotes additional parameters. The objective of $F$ is to extract user preferences, relationships, and global information by analyzing user-item interaction data, thereby enhancing recommendation performance. This function provides personalized recommendations by predicting the probability of user-item interactions, thereby increasing user satisfaction.

\subsection{Recommendation task optimization}
In graph-based contrastive learning methods, the pairwise ranking loss is one of the most common and frequently used approaches, and the formula is as follows:
\begin{flalign}
    &&
    L_{rec}=\sum_{u=0}^{M-1}  \sum_{(m,n)\in D_{u}}-log\sigma (\hat{r}_{um}-\hat{r}_{un})+\lambda \left \| E^{0}  \right \|^{2}
    &&
\end{flalign}
here, $\sigma(\cdot)$ denotes the sigmoid activation function, and $\lambda$ represents the regularization coefficient. $D_u = \left\{ (m, n) \mid m \in R_u \notag \right. \\ \left. \cap n \notin R_u \right\}$ denotes the training data containing positive and negative samples for user $u$, where $R_u$ is the set of items interacted with by user $u$.

\subsection{Graph contrastive learning}
In contrastive learning-based recommendation methods, the contrastive learning task is typically used as an auxiliary task. It leverages self-supervised signals to aid the main task in refining the embedding representations and operates within a multi-task framework:
\begin{flalign}
    &&
    L=L_{rec}+\alpha L_{cl}
    &&
\end{flalign}
here, $\alpha$ is a hyperparameter of the contrastive learning task, responsible for adjusting the participation ratio of contrastive learning and controlling the influence on the main task. $L_{cl}$ typically employs the InfoNCE loss function in most graph contrastive learning tasks:
\begin{flalign}
    &&
    L_{cl}=\sum_{i=0}^{I}-log\frac{exp(e_{i}^{\prime},e_{i}^{\prime \prime}/\tau )}{\sum_{j=0}^{I}exp(e_{i}^{\prime},e_{j}^{\prime \prime}/\tau)}
    &&
\label{Eq.4}
\end{flalign}
here, $I$ represents the batch, and $\tau$ denotes the temperature coefficient, both of which are common practices in InfoNCE. $e_{i}^{\prime}$ and $e_{i}^{\prime \prime}$ represent the embedding of the primary view and the augmented view of node $i$, respectively, which are the two contrastive objectives in the contrastive learning. These objectives are the same as those of node $j$. This loss function is a core component of the contrastive learning task.

\subsection{Two data augmentation methods}
\begin{figure}[ht]
\centering
\includegraphics[width=0.8\linewidth]{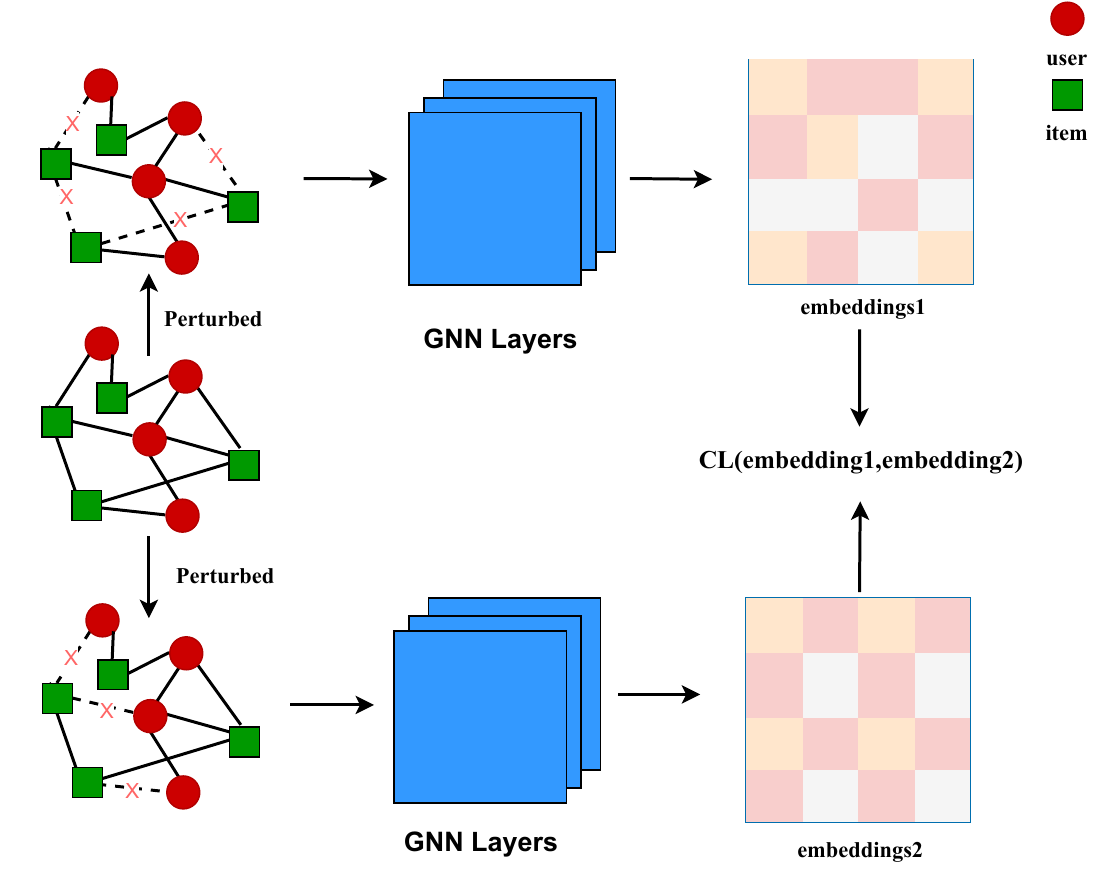}
\caption{GCL frameworks with structural augmentation}
\label{Fig.1}
\end{figure}

\begin{figure}[ht]
\centering
\includegraphics[width=0.9\linewidth]{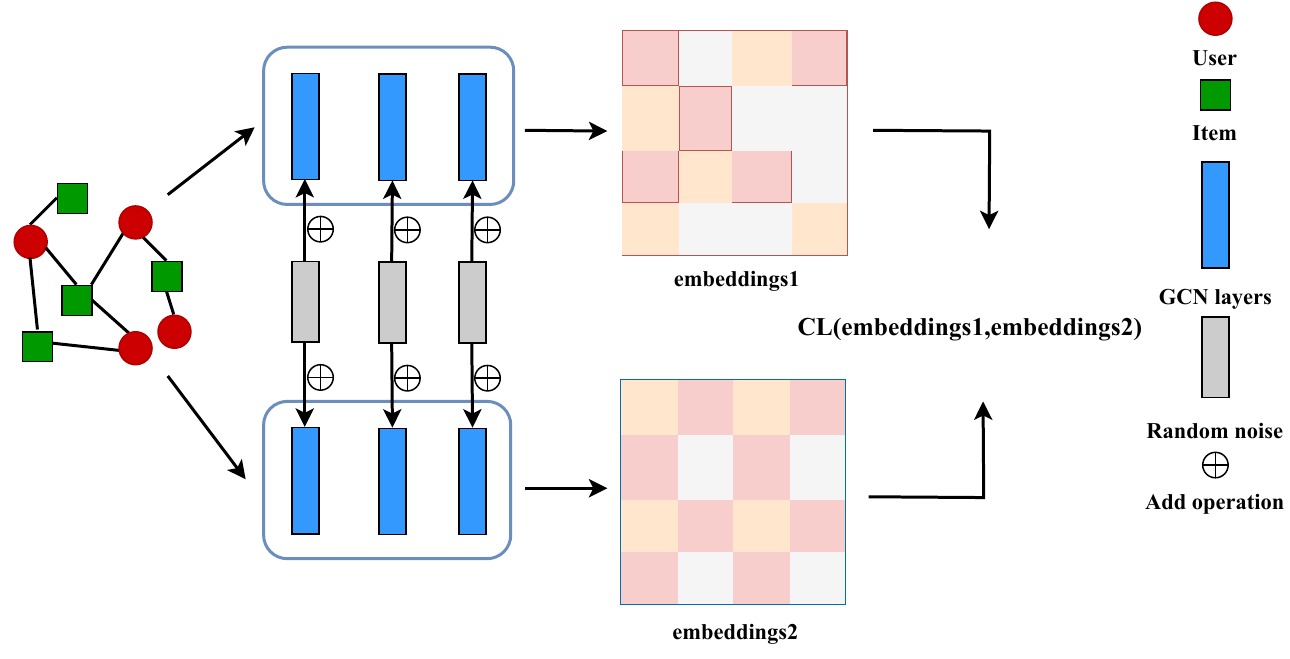}
\caption{GCL frameworks with feature augmentation}
\label{Fig.2}
\end{figure}

In this section, we introduce two common graph augmentation methods used in contrastive learning \cite{yang2023generative}. The first method is the structural augmentation approach, demonstrated by randomly deleting edges, as shown in Fig.\ref{Fig.1}. The second method is the feature augmentation approach, exemplified by adding random noise, as illustrated in Fig.\ref{Fig.2}.

Structural augmentation methods typically involve randomly removing or adding nodes or edges. As illustrated in Fig.\ref{Fig.1}, we generate two new graphs $G'$ and $G''$ by randomly deleting edges, learning their representation embeddings through a GNN, and computing the contrastive loss using the embeddings of these two graphs. The formula for the representation embeddings is as follows:
\begin{flalign}
    &&
    E^{\prime } =\eta (G^{\prime},E^{0}),\quad E^{\prime \prime }=\eta (G^{\prime \prime },E^{0})
    &&
\end{flalign}
here, $\eta(\cdot)$ represents the graph encoder used for generating perturbed graphs. However, this approach may exacerbate the data sparsity problem in recommendation tasks, as discarding part of the already sparse information might lead to the loss of certain critical information, reducing the key information obtained by the method and impairing its effectiveness.

Feature augmentation methods typically involve adding perturbations during the process of feature representation generation. As shown in Fig.\ref{Fig.2}, we introduce random noise during the GNN's aggregation of node information, thereby generating different embeddings, which are then subjected to contrastive learning:
\begin{flalign}
    &&
    E^{\prime } =\eta (E^{0},\epsilon \rho^{\prime }),E^{\prime \prime }=\eta (E^{0},\quad \epsilon \rho^{\prime \prime})
    &&
\end{flalign}
here, $\rho^{\prime}$ and $\rho^{\prime\prime}$ represent the noise, which is most commonly uniform noise, i.e., $\rho^{\prime}, \rho^{\prime\prime} \sim U(0,1)$. The parameter $\epsilon$ denotes the noise control parameter, used to regulate the magnitude of the noise and thereby control the extent of perturbation applied to the embeddings.

\section{Proposed method}

\begin{figure}[ht]
\centering
\includegraphics[width=\linewidth]{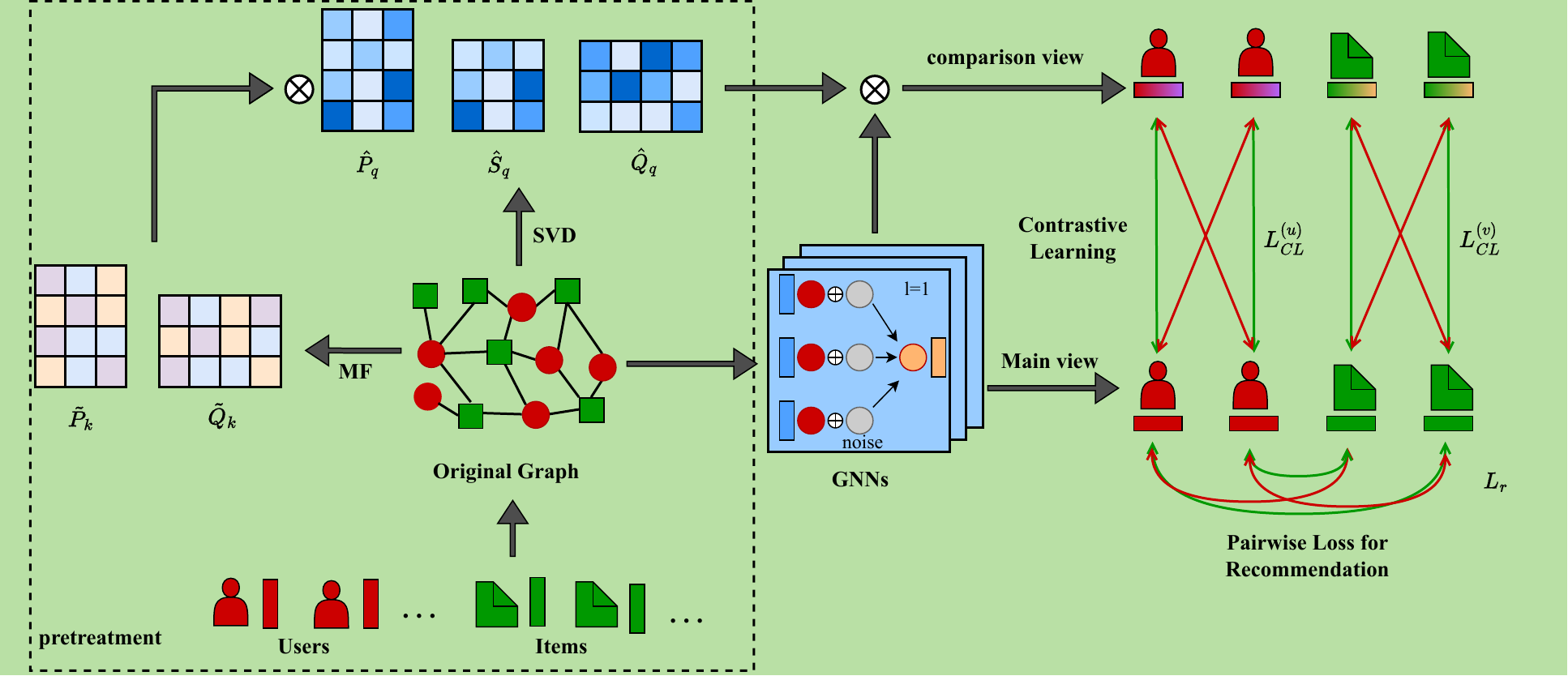}
\caption{Overall structure of HMFGCL}
\label{Fig.3}
\end{figure}

In this section, we elaborate on the implementation details of our proposed method, HMFGCL. As shown in Fig.\ref{Fig.3}, our approach is mainly divided into three parts. The first part is the data augmentation preprocessing module guided by MF and SVD, aiming to complementarily acquire global collaborative information. The second part is the GNN backbone, which involves the addition of noise and is designed to capture local user-item dependencies. The third part is the contrastive learning module, encompassing the fusion of local dependencies and global collaborative signals as well as the computation of contrastive loss.

\subsection{Architecture overview}
The information used to construct the rating matrix $A$, which represents user-item interactions, is extracted from the user-item interaction graph $G$. We normalize this matrix and convert it into a sparse matrix, which serves as our fundamental data.

First, in the data augmentation preprocessing module, we perform two separate factorization processes on the rating matrix $A$, namely, low-rank factorization guided by MF and SVD. The user latent feature matrices are denoted as $\tilde{P}_k$ and $\hat{P}_q$; the item latent feature matrices are represented as $\tilde{Q}_k$ and $\hat{Q}_q$, where $k$ is the dimensionality of the latent vectors in MF, and $q$ is the number of largest singular values in SVD. Both factorization approaches preserve significant global information, preparing for subsequent complementary fusion.

Next, in the GNN backbone, we introduce mixed noise $\Delta$ in the aggregation layer to induce feature perturbation, enhance the uniformity of representations, and optimize the embedding quality. Through multiple aggregation layers, we iteratively update the initial embeddings $E_{0}^{u}$and $E_{0}^{v}$, as well as the rating matrix $A$, to obtain the final user embeddings $E^{(u)}$ and item embeddings $E^{(v)}$.

Finally, we come to the contrastive learning module. The first step involves the integration of local dependencies with global collaborative information to form contrastive views. We merge the latent feature matrices $\tilde{P}_k$ and $\tilde{Q}_k$ generated by MF with the latent feature matrices $\hat{P}_q$ and $\hat{Q}_q$ obtained from SVD factorization as auxiliary information. These auxiliary information is then combined with the aggregation results of each layer to derive user representation embeddings $G^{(u)}$ and item representation embeddings $G^{(v)}$ that incorporate global collaborative information, forming contrastive views. The second step is the computation of the contrastive loss. We use the final aggregation results of the graph neural network, $E^{(u)}$ and $E^{(v)}$, as the main views for the contrastive learning task, and contrast them with the contrastive views that include global collaborative information, $G^{(u)}$ and $G^{(v)}$, to calculate the InfoNCE contrastive loss. Lastly, we use this contrastive loss to correct the main task of recommendation using the main views.

\subsection{Data augmentation preprocessing module}
We introduce global collaborative information through preprocessing and maximize the extraction of this information via complementary acquisition methods with MF and SVD, preparing to create contrastive views. We perform two factorization processes on the normalized adjacency matrix $A$. Using MF, we decompose it into a low-rank approximation matrix $\tilde{A} = \tilde{P} \tilde{Q}^{\top}$, where $\tilde{P} \in \mathbb{R}^{m \times k}$ represents the user latent factor matrix, $\tilde{Q} \in \mathbb{R}^{n \times k}$ represents the item latent factor matrix, and $k$ denotes the dimensionality of the latent vectors. Similarly, we use SVD to decompose the normalized adjacency matrix $A$ into an approximation matrix $\hat{A} = \hat{P} \hat{Q}^{\top}$, where $\hat{P} \in \mathbb{R}^{m \times q}$ represents the user latent factor matrix, $\hat{Q} \in \mathbb{R}^{n \times q}$ represents the item latent factor matrix, and $q$ denotes the number of largest singular values. The equations are as follows:

\begin{flalign}
    &&
    \tilde{P}_{k},\tilde{Q} _{k} = MF(A,k),
    \tilde{A}_{MF} = \tilde{P}_{k}\tilde{Q}_{k}^{\top}
    &&
\label{Eq.7}
\end{flalign}
\begin{flalign}
    &&
    \hat{P}_{q},\hat{s}_{q},\hat{Q} _{q} = SVD(A,q),
    \hat{A}_{SVD} = \hat{P}_{q}\hat{S}_{q}\hat{Q}_{q}^{\top}
    &&
\label{Eq.8}
\end{flalign}

The rationale for employing both factorization methods is twofold: Firstly, both MF and SVD can preserve significant user-item interactions, i.e., global collaborative information, through low-rank decomposition. Secondly, by leveraging the complementary relationship formed by combining these two methods, we maximize global information acquisition. This approach enhances the capture of user preferences and overall information gathering.

\subsection{The GNN backbone}
The core part of the model primarily generates the main views for both the recommendation task and the contrastive learning task. Generally, graph-based recommendation methods utilize multiple aggregation layers to gather information from neighboring nodes, thereby forming new node embeddings. Assuming the initial node embedding to be $E_{0}$, the update process through a graph neural network is defined as follows:
\begin{flalign}
    &&
    E_{l}=D^{-\frac{1}{2}}AD^{-\frac{1}{2}}E_{l-1} 
    &&
\label{Eq.9}
\end{flalign}
where $D$ is the degree matrix, recording the number of edges associated with each node. $E^{l}$ and $E^{l-1}$ represent the node embeddings for the $l$-th and $l-1$-th layers, respectively. When the number of graph convolution layers is $L$, the final node embeddings are read as follows:
\begin{flalign}
    &&
    E=Out(E_{0},E_{1},...,E_{L})
    &&
\label{Eq.10}
\end{flalign}
here, $Out$ does not refer to a specific method, as there are typically multiple ways to handle the aggregation results from multiple layers, such as summing the layers or taking the result of the last layer as the final output.

Building upon this foundation, we present our implementation of the GNN: First, we assign an embedding vector $e_m^{(u)}$ and $e_n^{(v)} \in \mathbb{R}^d$ to each user $u_m$ and each item $v_n$, where $d$ denotes the embedding size, which is a common practice in recommendation systems. Then, we define $E^{(u)} \in \mathbb{R}^{M \times d}$ and $E^{(v)} \in \mathbb{R}^{N \times d}$ as the sets of all users and items, respectively, where $M$ and $N$ represent the number of users and items. Finally, we introduce mixed noise interference embeddings, formulated as follows:
\begin{flalign}
    &&
    e_{m}^{\prime } = e_{m} + \Delta _{m}, \quad
    e_{n}^{\prime } = e_{n} + \Delta _{n}
    &&
\label{Eq.11}
\end{flalign}
where the noise vector $\Delta_m$ satisfies $\left| \Delta_m \right|^2 = \epsilon$ and $\epsilon$ is a small constant. The vector $\Delta_n$ operates similarly, and both $\Delta_m$ and $\Delta_n$ have dimensions identical to $e_m$ and $e_n$. Additionally, the noise vector is formulated as follows:
\begin{flalign}
    &&
    \Delta = \omega \odot (l_{1}\times  N_{u}  + l_{2}\times N_{g}),
    \omega \in \mathbb{R} ^{d}\sim U(0,1)
    &&
\end{flalign}
where $l_1$ and $l_2$ are two hyperparameters controlling the proportions of the two noises, with $l_1 + l_2 = 1$. $N_{g}$ and $N_{u}$ are Gaussian noise and uniform noise, respectively, and satisfy:
\begin{flalign}
    &&
    N_g \sim \mathcal{N}(0, 1),\quad N_u \sim \text{Uniform}(0, 1)
    &&
\end{flalign}

The purpose of this approach is to use lightweight noise perturbations to enhance the uniformity of the embeddings, thereby improving the representation quality. Based on Eq.\ref{Eq.9} and Eq.\ref{Eq.11}, we rewrite the GNN aggregation process as follows:
\begin{flalign}
    &&
    E_{l}=D^{-\frac{1}{2}}AD^{-\frac{1}{2}}E^{\prime}_{l-1} 
    &&
\end{flalign}
here, $E_{l-1}'$ represents the embedding representation at layer $l-1$ after adding noise, specifically for user $m$ and item $n$ at layer $l$:
\begin{flalign}
    &&
    z_{m,l}^{(u)} =\sigma (A_{m,:} \cdot E_{l-1}^{(v)'}), \quad
    z_{n,l}^{(v)} =\sigma (A_{:,n} \cdot E_{l-1}^{(u)'})
    &&
\end{flalign}
here, $z_{m,l}^{(u)}$ and $z_{n,l}^{(v)}$ denote the aggregated embeddings for user $m$ and item $n$ at layer $l$. The function $\sigma(\cdot)$ represents the activation function, and $A$ is the normalized adjacency matrix. $E_{l-1}^{(v)'}$ and $E_{l-1}^{(u)'}$ represent the item embeddings and user embeddings at layer $l-1$ after noise has been added, respectively.

\subsection{Graph contrastive learning}
Firstly, we fuse the global information \(\tilde{A}_{MF}\) and \(\hat{A}_{SVD}\) obtained from the preprocessing data augmentation module with local dependencies to form the contrastive view embeddings. From Eq.\ref{Eq.7} and Eq.\ref{Eq.8}, we obtain:
\begin{flalign}
    G_{l}^{(u)}&=\sigma(\tilde{A}_{MF}E_{l-1}^{(v)}\hat{A}_{SVD})\\
    &=\sigma(\tilde{P}_{k}\tilde{Q}_{k}^{\top}E_{l-1}^{(v)}\hat{P}_{q}\hat{S}_{q}\hat{Q}_{q}^{\top})
\label{Eq.16}
\end{flalign}
\begin{flalign}
    G_{l}^{(v)}&=\sigma(\tilde{A}_{MF}^{\top}E_{l-1}^{(u)}\hat{A}_{SVD}^{\top})\\
    &=\sigma(\tilde{Q}_{k}\tilde{P}_{k}^{\top}E_{l-1}^{(u)}\hat{Q}_{q}\hat{S}_{q}\hat{P}_{q}^{\top})
\label{Eq.17}
\end{flalign}
here, $G_{l}^{(u)}$ and $G_{l}^{(v)}$denote the user and item embeddings that are fused with global signals, representing the contrastive views that incorporate global collaborative information. From Eq.\ref{Eq.16} and Eq.\ref{Eq.17}, we obtain the information fusion process for user $m$ and item $n$ at layer $l$:
\begin{flalign}
    &g_{m,l}^{(u)} = \sigma (\tilde{A_{m,:} } \cdot E_{l-1}^{(v)} \cdot \hat{A_{m,:}}),\\
    &g_{n,l}^{(v)} = \sigma (\tilde{A_{:,n} } \cdot E_{l-1}^{(u)} \cdot \hat{A_{:,n}})
\end{flalign}
The $G_{l}^{(u)}$ and $G_{l}^{(v)}$ obtained from Eq.\ref{Eq.16} and Eq.\ref{Eq.17} correspond to the contrastive view embeddings.

Next is the calculation of contrastive loss. Traditional GCL methods typically adopt a three-view paradigm, where two additional views are generated through data augmentation and contrasted for learning, while the main view does not directly participate in the InfoNCE calculation. However, this paradigm leads to high methodological complexity, as both views need to be learned through graph neural networks. In our approach, we abandon the traditional three-view structure. The embeddings $E^{(u)}$ and $E^{(v)}$ generated by the main view serve as one part of the InfoNCE calculation, while the contrastive view embeddings $G^{(u)}$ and $G^{(v)}$ constitute the other part. According to Eq.\ref{Eq.4}, for user $m$ and item $n$, we rewrite the formula as follows:
\begin{flalign}
    &&
    L_{CL}^{(u)}=\sum_{m=0}^{M} \sum_{l=0}^{L} - log\frac{exp(s(z_{m,l}^{(u)},g_{m,l}^{(u)}/\tau ))}{{\textstyle \sum_{m^{\prime}=0}^{M}} exp(s(z_{m,l}^{(u)},g_{m^{\prime } ,l}^{(u)}/\tau ))}
    &&
\end{flalign}
\begin{flalign}
    &&
    L_{CL}^{(v)}=\sum_{n=0}^{N} \sum_{l=0}^{L} - log\frac{exp(s(z_{n,l}^{(v)},g_{n,l}^{(v)}/\tau ))}{{\textstyle \sum_{n^{\prime}=0}^{N}} exp(s(z_{n,l}^{(v)},g_{n^{\prime } ,l}^{(v)}/\tau ))}
    &&
\end{flalign}
here, $s(\cdot)$ denotes the cosine similarity, and $\tau$ represents the temperature. In Eq. \ref{Eq.21} and Eq. \ref{Eq.22}, we elaborate on the concurrent optimization of the contrastive learning loss and the adjustment of the primary objective function for the recommendation task. Specifically, $\hat{y}_{m,p_s}$ and $\hat{y}_{m,n_s}$ represent the predicted scores of the positive and negative item pairs for user $m$.
\begin{flalign}
    L=L_{r}+\lambda_{1}\cdot(L_{s}^{(u)}+L_{s}^{(v)})+\lambda_{2}\cdot\left \| \Theta  \right \| _{2}^{2}
\label{Eq.21}
\end{flalign}

\begin{flalign}
    L_{r}=\sum_{m=0}^{M} \sum_{s=1}^{S} max(0,1-\hat{y}_{m,p_{s}}+\hat{y}_{m,n_{s}})
\label{Eq.22}
\end{flalign}

\subsection{Algorithm implementation}
Algorithm \ref{alg:HMFGCL} describes the implementation process of the proposed method. The inputs include the user-item interaction matrix $A$, the graph $G$, the low-rank approximation matrices $\tilde{A}$ and $\hat{A}$, the user embeddings $E^{(u)}$, the item embeddings $E^{(v)}$, the training parameters ${{\bf W}_i, {\bf b}_i}_{i=1}^L$, ${{\bf u}}_{u \in U}$, and ${{\bf v}}_{v \in V}$, as well as the hyperparameters $GNN(\cdot)$, $f(\cdot)$, $Frobenius(\cdot)$, and $fusion(\cdot)$. The output is the prediction function $F(u, v)$.

The algorithm first defines two functions. The first is MF, which aims to capture the global information of users and items for subsequent feature fusion. The main input parameters are the user-item interaction matrix $A$, the latent vector dimension $k$, and the number of iterations for optimization $T$. This function computes the Frobenius norm as the loss function and employs simple gradient descent to optimize it, resulting in the low-rank approximation matrix $\tilde{A}$. Upon completion, the function returns the latent feature matrices $\tilde{P}_{k}$ and $\tilde{Q}_{k}$. The second function is SVD, which serves a purpose similar to MF. It obtains global information about users and items through truncated singular value decomposition, preparing the data for later feature fusion. The primary parameters are the user-item interaction matrix $A$ and the number of largest singular values $q$. The core of this function is the use of the torch.svd\_lowrank() function from the torch library within the Pytorch framework to obtain the approximate matrix $\hat{A}$. The function returns the approximate feature matrices $\hat{P}_{q}$, $\tilde{S}_{q}$, and $\hat{Q}_{q}$.

The main loop of the algorithm iteratively updates the parameters until convergence is reached for each batch of users $u$ and items $v$. First, we utilize the MF and SVD functions to obtain the latent feature matrices of users, $\tilde{P}_{k}$ and $\hat{P}_{q}$, as well as the latent feature matrices of items, $\tilde{Q}_{k}$ and $\hat{Q}_{q}$. Subsequently, we employ the GNN to aggregate the adjacency matrix $\tilde{A}$ with the node feature representations $E$ from the previous layer and introduce noise to obtain the user representations $z^{(u)}$ and item representations $z^{(v)}$.

Next, the user embedding representation $z^{(u)}$ is fused with the latent feature matrices obtained from the MF function, $\tilde{P}_{k}$, and the latent feature matrices obtained from the SVD function, $\hat{P}_{q}$ and $\hat{S}_{q}$, to derive the contrastive view representation embedding of the user, denoted as $g^{(u)}$. Similarly, the item embedding representation $z^{(v)}$ is fused with the latent feature matrices obtained from the MF function, $\tilde{Q}_{k}$, and the latent feature matrices from the SVD function, $\hat{Q}_{q}$ and $\hat{S}_{q}$, resulting in the contrastive view representation embedding of the item, denoted as $g^{(v)}$. The fusion process is represented by the function $fusion(\cdot)$. The representations $g^{(u)}$ and $g^{(v)}$ thus reflect the user and item embeddings enriched with global collaborative information.

\SetKwProg{Fn}{Function}{}{}
\begin{algorithm}[!h]
    \caption{HMFGCL}
    \label{alg:HMFGCL}
    \KwIn{User-item interaction matrix $A$; User-item bipartite graph $G$;
           MF approximation matrix $\tilde{A}$; SVD approximation matrix $\hat{A}$;
           User embeddings $E^{(u)}$; Item embeddings $E^{(v)}$;Additional parameters $\Pi$ ; \\ \quad \quad \quad
          Train parameters: $\{{\bf W}_i, {\bf b}_i\}_{i=1}^L$,$\{{\bf u}\}_{u \in {U}}$, 
          $\{{\bf v}\}_{v \in {V}}$; \\ \quad \quad \quad
          Hyper-parameters: $GNN(\cdot)$,$f(\cdot)$,$Frobenius(\cdot)$, $fusion(\cdot)$ }
    \KwOut{Prediction function $F(u,v\mid G, A, \tilde{A} ,\hat{A},\Pi )$}
    \BlankLine
    \Fn{\rm{MF} $(A, k, T)$}{
        $m,n=shape(\tilde{A})$\\
        $\tilde{P}_{k} = Randn(m,k)$, $\tilde{Q}_{k} = Rnadn(n,k)$ \\
        \For{$t$ in $T$}{
            $\tilde{A}=\tilde{P}_{k}\tilde{Q}_{k}^{\top}$\\
            $loss=Frobenius(A,\tilde{A})$\\
            loss.backward()\\
            optimizer.step()\\
        }
        \Return $\tilde{P}_{k}$, $\tilde{Q}_{k}$\;
    }
    \Fn{\rm{SVD} $(A, q)$}{
        $m,n=shape(\hat{A})$\\
        $\hat{P}_{q} = Randn(m,q)$, $\hat{Q}_{q} = Rnadn(n,q)$, $\hat{S}_{q} = Rnadn(q,q)$ \\
        $\hat{P}_{q}$,$\hat{S}_{q}$, $\hat{Q}_{q} = svd\_lowrank(A,q)$\\
        \Return $\hat{P}_{q}$,$\hat{S}_{q}$, $\hat{Q}_{q}$\;
    }
    \While{HMFGCL not converge}{
        \For{$u, v$ in $A$}{
            $\tilde{P}_{k}, \tilde{Q}_{k} \leftarrow $ MF $(A, k, T)$\;
            $\hat{P}_{q},\hat{Q}_{q},\hat{Q}_{q} \leftarrow $ SVD $(A, q, T)$\;
            $z^{(u)} \leftarrow GNN(E^{(u)},A)$\;
            $z^{(v)} \leftarrow GNN(E^{(u)},A)$\;
            $g^{(u)} \leftarrow fusion(\tilde{P}_{k},z^{(u)},\hat{P}_{q},\hat{S}_{q})$\;
            $g^{(v)} \leftarrow fusion(\tilde{Q}_{k},z^{(v)},\hat{Q}_{q},\hat{S}_{q})$\;
            Compute loss $L_{CL}^{(u)}=InfoNCE (z^{(u)}, g^{(u)})$\;
            Compute loss $L_{CL}^{(v)}=InfoNCE (z^{(v)}, g^{(v)})$\;
            Compute the predicted label $\hat y_{uv} = f({z^{(u)}}, {z^{(v}})$\;
            Update the parameters using gradient descent optimization.
        }
    }
    \Return$F$\;
    \BlankLine
\end{algorithm}

Subsequently, we compute the InfoNCE loss function for the auxiliary task to aid the main task in refining the representation embeddings. We utilize the primary view user and item embeddings $z^{(u)}$ and $z^{(v)}$, along with the contrastive view embeddings $g^{(u)}$ and $g^{(v)}$, which incorporate global collaborative information, as the basis for calculating the InfoNCE loss. Specifically, we compute the user contrastive loss $L_{CL}^{(u)}$ and the item contrastive loss $L_{CL}^{(v)}$. Finally, the function $f(\cdot)$ is employed to complete the prediction of user-item interactions.

Finally, we iteratively refine the parameters through gradient descent until convergence, yielding the prediction function $F$.

We estimated the time complexity of the algorithm by integrating the above steps, assuming a total of $I$ user-item interactions.
\begin{flalign}
    &&
    O(C_{MF}+C_{SVD}+N\times (C_{GNN}+L\cdot C_{agg})+C_{grad})
    &&
\end{flalign}
here, $L$ denotes the depth of the network, $C_{MF}$ represents the computational complexity of the MF operation, $C_{SVD}$ represents the computational complexity of the SVD operation, $C_{agg}$ represents the computational complexity of the aggregation operation, $C_{GCN}$ represents the computational complexity of the message passing in GCN, and $C_{grad}$ represents the computational complexity of performing gradient descent to update parameters.

\section{Results and discussion}

To validate the efficacy of our proposed method, we conducted comprehensive experimental evaluations. In the first part, we introduce the datasets utilized in this study. In the second part, we present the methodologies employed for comparative analysis, including both classical and contemporary approaches. In the third part, we provide the results and analysis of our comparative experiments. In the fourth part, we explore the impact of hyperparameters on the performance of the methods through experimentation. In the fifth part, we analyze the convergence curves of the loss functions to investigate the influence of the loss function on the model. In the sixth part, we perform ablation studies to assess the effectiveness of each module in our proposed method. Finally, we provide a comprehensive exposition of our experiments.

\subsection{DataSets}
To ensure the credibility of our experimental results, we conducted experiments on three publicly available datasets: (1) MovieLens-100k (ML-100k), a classic recommendation system dataset maintained by GroupLens Research Lab, which includes user ratings of movies from the MovieLens website. (2) ModCloth, provided by the ModCloth website, contains a large number of clothing products and user reviews. It plays a crucial role in market trend analysis, user behavior research, and the development and evaluation of recommendation systems. (3) LastFM, a commonly used dataset in recommendation systems and music recommendation research, evaluates the performance of algorithms in scenarios involving user-item interactions and sparse data. This dataset primarily includes information on user listening records. Due to the limitations of our experimental hardware and the advantages of our method in handling small-scale datasets, we preprocessed the ModCloth and LastFM datasets by randomly selecting 1000 users and their related items, resulting in the datasets ModCloth-1000 and LastFM-1000, which were used for our experiments.

\subsection{Baseline}
We compare our method against other baseline approaches.
\begin{itemize}
    \item	\textbf{DMF} \cite{xue2017deep} is one of the most classic recommendation methods implemented based on matrix factorization.
    \item	\textbf{GCMC} \cite{berg2017graph} is a recommender algorithm that combines graph convolutional neural networks with matrix completion. It proposes a differentiable message-passing graph autoencoder framework for acquiring user and item representations.
    \item	\textbf{NGCF} \cite{wang2019neural} is a collaborative filtering algorithm that employs convolutional neural networks to model user-item interactions as a graph and compute embeddings for both users and items.
    \item   \textbf{EASE} \cite{steck2019embarrassingly} is based on user behavior sequences. It introduces a feature learning method called "Exchange-Only Model," which reduces model parameters and computational complexity.
    \item   \textbf{LightGCN} \cite{he2020lightgcn} is one of the widely referenced methods in graph neural network-based recommendation systems and is a lightweight recommendation approach. It utilizes multiple layers of Graph Convolutional Networks (GCN) to learn embeddings for both users and items.
    \item   \textbf{ENMF} \cite{chen2020efficient} is implemented using a simple neural matrix factorization framework, learning embedding representations from entity matrices, and employing a non-sampling strategy.
    \item   \textbf{SimpleX\cite{mao2021simplex}} demonstrates that the choice of loss function and negative sampling rate is equally important, consisting of a cosine contrastive loss (CCL) combined with a simple collaborative filtering model.
    \item   \textbf{SGL\cite{wu2021self}} applies self-supervised learning on the user-item graph, utilizing self-supervised tasks as auxiliary tasks to refine the representation learning of the main task.
    \item   \textbf{SimGCL\cite{yu2022graph}} uses uniform noise to create contrastive views, abandoning traditional graph augmentation methods in order to learn a more uniform representation distribution.
    \item   \textbf{LightGCL\cite{cai2023lightgcl}} is an advanced lightweight model that introduces a singular value decomposition-guided graph augmentation technique, integrating contextual information into user-item information to enhance graph contrastive learning.
    \item   \textbf{DirectAU\cite{wang2022towards}} differs from the majority of recommendation system research approaches by focusing on proving factors influencing collaborative filtering, proposing that both the alignment and uniformity of representations impact recommendation performance.
    \item   \textbf{XSimGCL\cite{yu2023xsimgcl}} is an improved version of SimGCL, eliminating redundant data augmentation operations to make the model more lightweight, and introducing cross-layer contrastive learning to reduce time costs.
\end{itemize}

\subsection{Comparative experiment}
Table~\ref{Tab.1} compares the performance of our proposed method with several baseline methods. The best results of the experiments are highlighted with gray shading, and we provide further explanation and analysis for these results. Our study uses $R@k$ (Recall) and $N@k$ (NDCG) to evaluate the recommendation results at $k=10$ and $k=20$.

\begin{table*}[htbp]
\centering
\caption{
We conducted a comprehensive performance comparison between our proposed method and the baseline method under clean training conditions across three publicly available datasets. The best results are indicated in bold, while the second-best results are underlined. The "\% Improve." metric represents the relative improvement of our proposed method compared to the best baseline result.}
\label{Tab.1}
\resizebox{\textwidth}{!}{
\begin{tabular}{l|cccc|cccc|cccc}
\hline
& \multicolumn{4}{c|}{\textbf{ML-100k}} & \multicolumn{4}{c|}{\textbf{ModCloth-1000}} & \multicolumn{4}{c}{\textbf{LastFM-1000}} \\
\textbf{Methods} & \textbf{R@10}  & \textbf{R@20}  & \textbf{N@10}  & \textbf{N@20}  & \textbf{R@10}  & \textbf{R@20}  & \textbf{N@10}  & \textbf{N@20} & \textbf{R@10}  & \textbf{R@20}  & \textbf{N@10}  & \textbf{N@20} \\ \hline
\textbf{DMF} & 0.1947 & 0.3060 & 0.3487 & 0.3552 & 0.1553 & 0.2549 & 0.0801 & 0.1052 & 0.0678 & 0.1023 & 0.0793 & 0.0981 \\
\textbf{GCMC} & 0.1639 & 0.2411 & 0.3041 & 0.2966 & 0.1567 & 0.2578 & 0.0819 & 0.1071 & 0.1068 & 0.1526 & 0.1274 & 0.1527 \\
\textbf{NGCF} & 0.1997 & 0.3121 & 0.1821 & 0.3637 & 0.1837 & 0.2269 & 0.0837 & 0.1043 & 0.1441 & 0.2157 & 0.1691 & 0.2081 \\
\textbf{EASE} & 0.1783 & 0.2486 & 0.2143 & 0.2527 & 0.2045 & 0.2924 & 0.1144 & 0.1362 & 0.1718 & 0.2426 & 0.2065 & 0.2457 \\
\textbf{LightGCN} & 0.1662 & 0.2507 & 0.3053 & 0.3012 & 0.1425 & 0.2142 & 0.0795 & 0.0981 & 0.167 & 0.2355 & 0.2027 & 0.2406 \\
\textbf{ENMF} & 0.1659 & 0.2563 & 0.2991 & 0.2989 & 0.1945 & 0.2877 & 0.1112 & 0.1349 & 0.1645 & 0.2318 & 0.1974 & 0.2341 \\
\textbf{SimpleX} & 0.1831 & 0.2863 & 0.3375 & 0.3381 & 0.2017 & 0.2579 & 0.1097 & 0.1245 & 0.1323 & 0.2024 & 0.1416 & 0.1804 \\
\textbf{SGL} & 0.2037 & 0.3131 & 0.3809 & 0.3776 & 0.061 & 0.1001 & 0.0305 & 0.0408 & 0.1719 & 0.2401 & 0.2075 & 0.245 \\
\textbf{SimGCL} & 0.1057 & 0.1705 & 0.1451 & 0.1618 & 0.0617 & 0.1142 & 0.0296 & 0.0426 & 0.032 & 0.0436 & 0.0382 & 0.0446 \\
\textbf{LightGCL} & \underline{0.2086} & \underline{0.3223} & \underline{0.3726} & \underline{0.3743} & \underline{0.2529} & \underline{0.2957} & \underline{0.1987} & \underline{0.2096} & \underline{0.1679} & \underline{0.2369} & \underline{0.2025} & \underline{0.2405} \\
\textbf{DirectAU} & 0.1209 & 0.1925 & 0.1712 & 0.189 & 0.0621 & 0.1159 & 0.0415 & 0.0553 & 0.0882 & 0.1404 & 0.0979 & 0.1266 \\
\textbf{XSimGCL} & 0.1346 & 0.2197 & 0.2088 & 0.2231 & 0.1262 & 0.1844 & 0.0821 & 0.0974 & 0.1635 & 0.2384 & 0.2006 & 0.2421 \\\hline
\cellcolor{gray!16}\textbf{HMFGCL} & \cellcolor{gray!16}\textbf{0.2232*} & \cellcolor{gray!16}\textbf{0.334*} & \cellcolor{gray!16}\textbf{0.4034*} & \cellcolor{gray!16}\textbf{0.4001*} & \cellcolor{gray!16}\textbf{0.2785*} & \cellcolor{gray!16}\textbf{0.3232*} & \cellcolor{gray!16}\textbf{0.2206*} & \cellcolor{gray!16}\textbf{0.2321*} & \cellcolor{gray!16}\textbf{0.1815*} & \cellcolor{gray!16}\textbf{0.2544*} & \cellcolor{gray!16}\textbf{0.2184*} & \cellcolor{gray!16}\textbf{0.2584*} \\

\textbf{\% Improve.} & 7.01\% & 3.63\% & 8.27\% & 6.89\% & 10.12\% & 9.30\% & 11.02\% & 10.73\% & 10.24\% & 7.39\% & 7.85\% & 7.44\% \\ \hline
\end{tabular}
}
\end{table*}

All our experiments are conducted using the RecBole\cite{zhao2021recbole} framework, an open-source platform specifically designed for rapid prototyping and evaluation of recommendation algorithms. The platform includes various recommendation tasks, such as traditional recommendation, sequential recommendation, and knowledge graph-based recommendation. It aims to advance research in the field of recommendation systems by providing an efficient development and evaluation platform. Featuring a comprehensive set of functionalities, the platform offers multiple recommendation algorithms and experimental tools for tasks such as dataset preprocessing, rapid hyperparameter tuning, and result analysis. All baseline methods presented in the table originate from the RecBole framework and utilize the optimal parameters provided by the framework.

For the hyperparameter settings, we configured the embedding size to be 64 and used an Adam optimizer with a learning rate of 0.001. The coefficient for L2 regularization was set to $10^{-5}$, and the aggregation layer was set to 2, which are commonly used parameters in previous studies. The coefficient for the self-supervised learning loss function was set to 0.003, while the noise coefficient was set to 0.1. The noise ratio was set to 8:2 (Gaussian noise: uniform noise), and the dimension for matrix factorization was set to 5. Further exploration experiments regarding individual hyperparameter settings will be elaborated in subsequent sections.

As shown in Table~\ref{Tab.1}, our method outperforms all baselines across three public datasets. Compared to the second-best baseline, LightGCL, the evaluation metrics R@10 and N@10 on the ml-100k dataset increased by 7.01\% and 8.27\%, respectively. The improvements for R@20 and N@20 were slightly lower, at 3.63\% and 6.89\%, respectively. On the ModCloth-1000 dataset, R@10 and N@10 saw increases of 10.12\% and 11.02\%, respectively, while R@20 and N@20 improved by 9.30\% and 10.73\%, respectively. On the LastFM dataset, R@10 and N@10 increased by 10.24\% and 7.85\%, respectively, and R@20 and N@20 improved by 7.39\% and 7.44\%, respectively. These results substantiate the advantages of our proposed method.

\subsection{Hyparameters analysis}
In this section, we adjust several hyperparameter values to investigate their impact on the performance of the proposed method. First, we vary the dimensionality of the latent vectors $k$ and the number of maximum singular values $q$ to observe how these factors affect the method's performance. Next, we alter the number of layers in the graph neural network to examine the effect of network depth on performance. Finally, we adjust the size of the embedding dimensions to investigate the impact of embedding size on the method.

\subsubsection{Impact of latent vector dimension}
In this section, we investigate the impact of the dimensionality of latent vectors on the performance of matrix factorization methods. As shown in Table \ref{Tab.2}, we conduct experiments on both the ML-100K and LastFM-1000 datasets to observe the variation in method performance. The data reveals that the method achieves its highest performance when the dimensionality of the latent vectors is set to 5. As the dimensionality gradually increases, there is a slight decline in the method's performance.
\begin{table}[htbp]
\centering
\caption{We conducted experiments on two datasets, using Recall and NDCG as evaluation metrics, with k values set to 10 and 20, denoted as R@10, R@20, N@10, and N@20.}
\label{Tab.2}
\resizebox{0.5\textwidth}{!}{
\begin{tabular}{l|cccc|cccc}
\hline
& \multicolumn{4}{c|}{\textbf{ML-100K}} & \multicolumn{4}{c}{\textbf{LastFM-1000}} \\
\textbf{Dimension} & \textbf{R@10}  & \textbf{R@20}  & \textbf{N@10}  & \textbf{N@20}  & \textbf{R@10}  & \textbf{R@20}  & \textbf{N@10}  & \textbf{N@20} \\ \hline
\textbf{5} & 0.2232 & 0.3351 & 0.4034 & 0.4008 & 0.1815 & 0.2544 & 0.2184 & 0.2584 \\
\textbf{10} & 0.2189 & 0.3322 & 0.3973 & 0.3963 & 0.1716 & 0.2419 & 0.2092 & 0.248 \\
\textbf{15} & 0.2201 & 0.3296 & 0.399 & 0.3959 & 0.1719 & 0.2434 & 0.207 & 0.2459 \\
\textbf{20} & 0.2227 & 0.3319 & 0.4032 & 0.3977 & 0.1701 & 0.2438 & 0.2057 & 0.2459 \\ \hline
\end{tabular}
}
\end{table}
From the experimental results in Table \ref{Tab.2}, we can observe that the method achieves its maximum performance when the dimensionality is 5. On the ML-100K dataset, when the dimensionality is 5, R@10 and R@20 are 0.2232 and 0.3351, respectively, while N@10 and N@20 are 0.4034 and 0.4008, respectively. However, as the dimensionality increases to 10, 15, and 20, the method's performance exhibits varying degrees of decline. The performance drop is most significant when the dimensionality is 10, with R@10 and R@20 being 0.2189 and 0.3322, respectively, and N@10 and N@20 being 0.3973 and 0.3963, respectively.

On the LastFM-1000 dataset, we observe a similar pattern where the method performs best when the dimensionality is 5. Specifically, R@10 and R@20 are 0.1815 and 0.2544, respectively, while N@10 and N@20 are 0.2184 and 0.2584, respectively. As the dimensionality increases, the method's performance declines, with the worst performance occurring when the dimensionality is 20. At this dimensionality, R@10 and R@20 are 0.1701 and 0.2438, respectively, and N@10 and N@20 are 0.2057 and 0.2459, respectively. This suggests that smaller latent vector dimensions in matrix factorization are effective in our method, enabling the capture of latent features of users and items. Conversely, larger latent vector dimensions may harm the method's performance, possibly due to an increase in redundant information as the dimensionality grows, leading to more noise and thereby reducing the method's effectiveness.

Experiments conducted on both datasets indicate that the dimensionality of latent vectors in matrix factorization is crucial for our method, as different choices of latent vector dimensions can yield varying impacts on performance. Therefore, selecting an appropriate dimensionality for the latent vectors is essential, as excessively large dimensions can significantly degrade the performance of the method.

\subsubsection{Impact of the number of maximum singular values}
\begin{figure}[ht]
    \centering
    \includegraphics[width=0.8\linewidth]{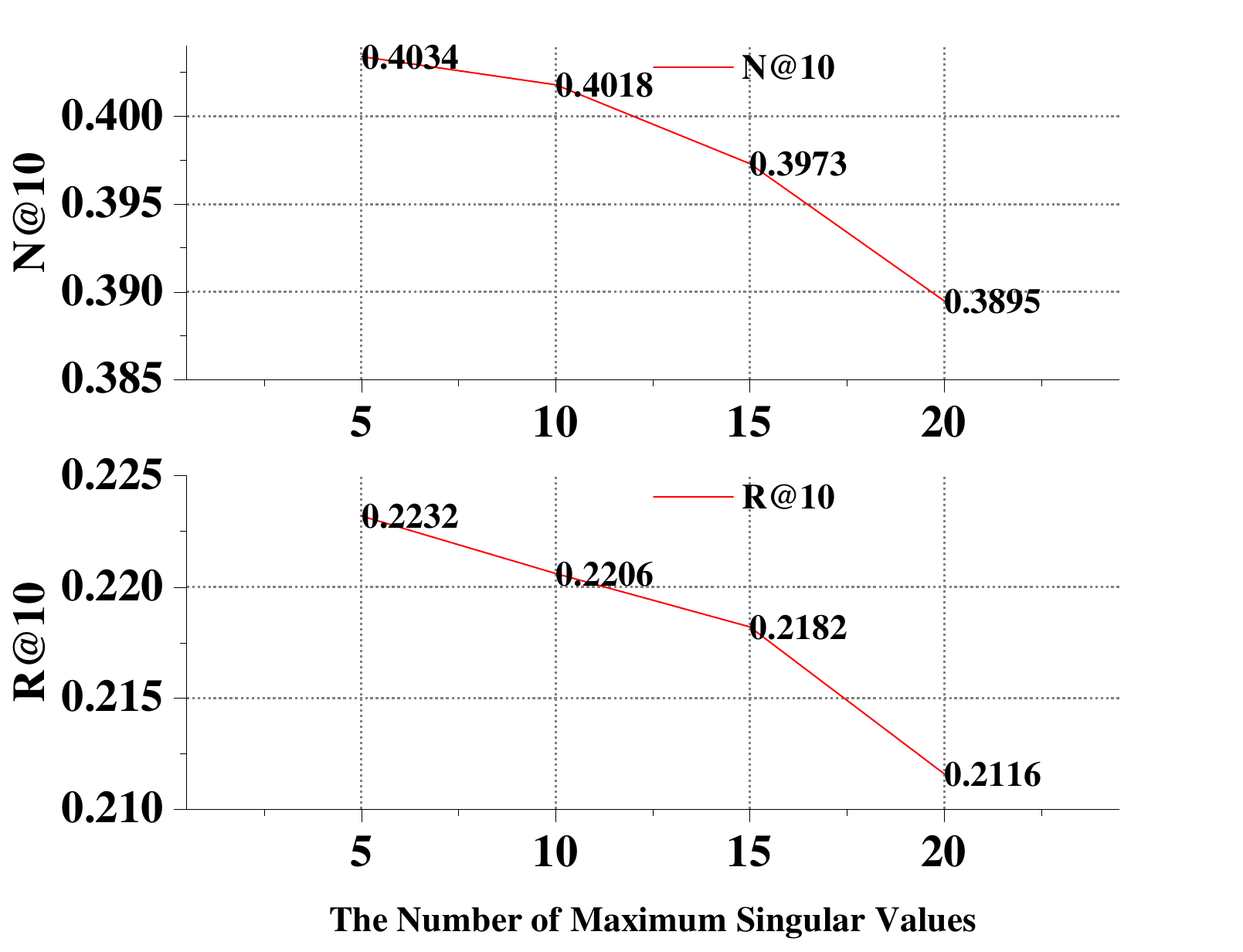}
    \caption{Impact of the number of maximum singular values(ML-100K) with R@10 and N@10}
    \label{Fig.4}
\end{figure}

\begin{figure}[ht]
    \centering
    \includegraphics[width=0.8\linewidth]{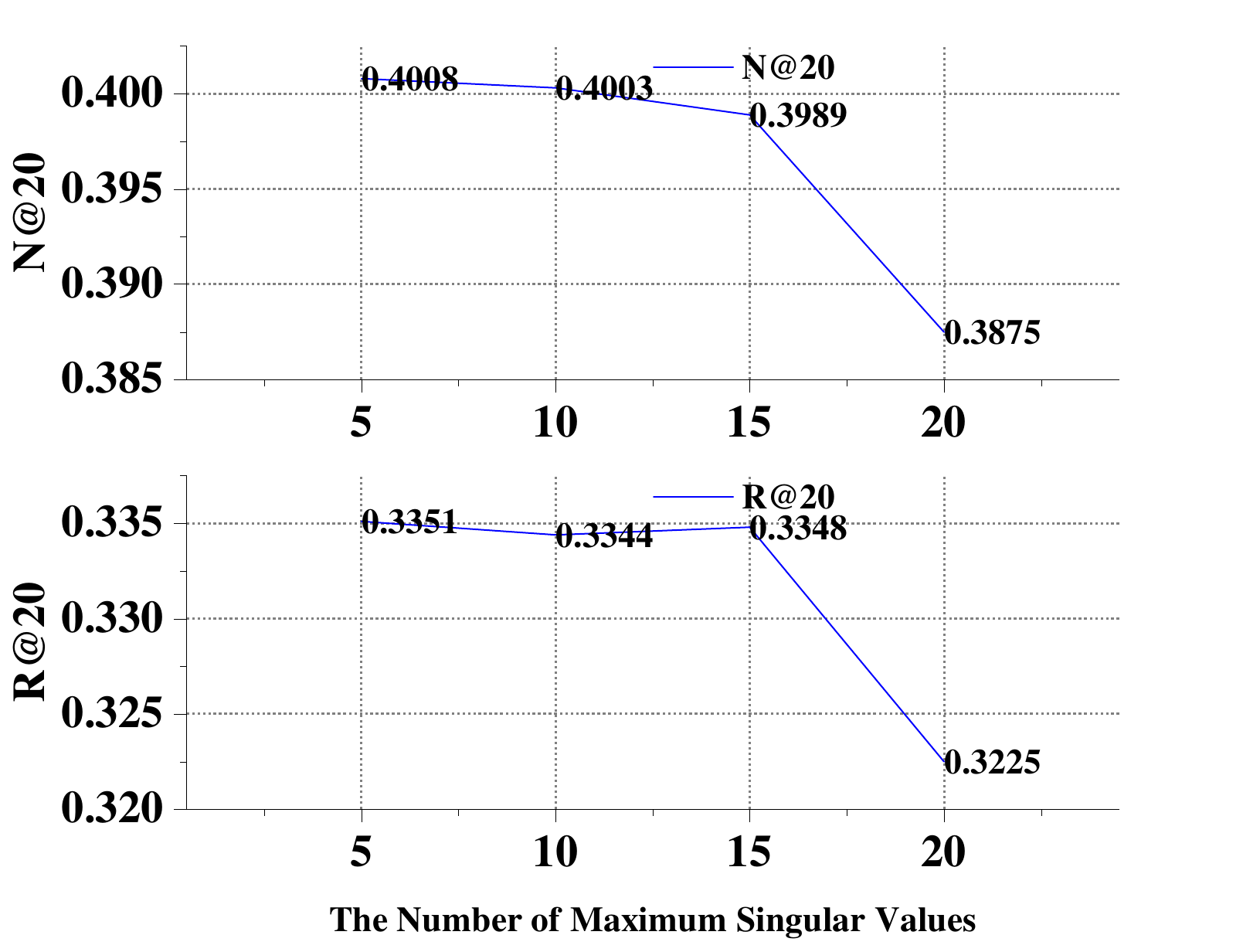}
    \caption{Impact of the number of maximum singular values(ML-100K) with R@20 and N@20}
    \label{Fig.5}
\end{figure}

\begin{figure}[ht]
    \centering
    \includegraphics[width=0.8\linewidth]{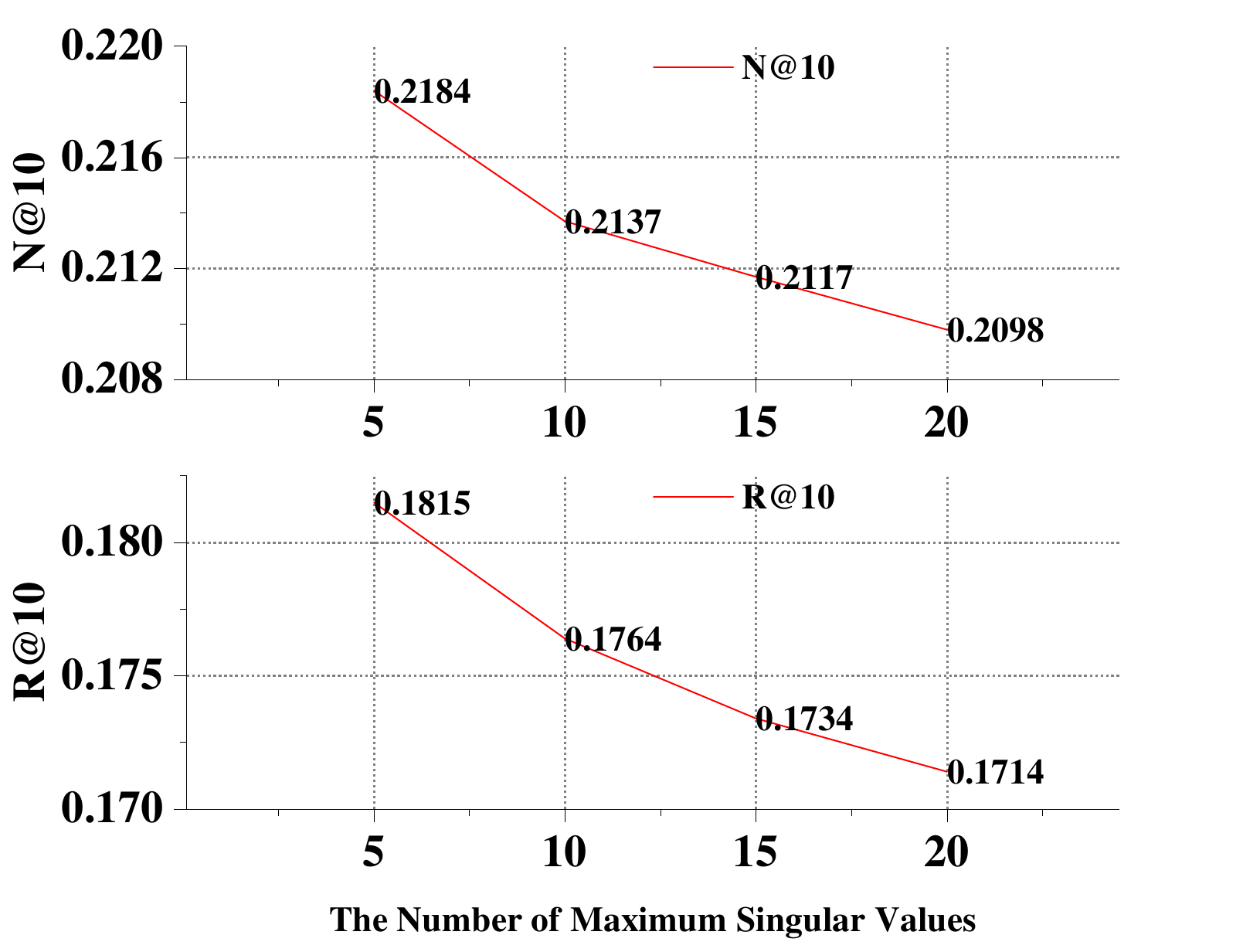}
    \caption{Impact of the number of maximum singular values(LastFM-1000) with R@10 and N@10}
    \label{Fig.6}
\end{figure}

\begin{figure}[ht]
    \centering
    \includegraphics[width=0.8\linewidth]{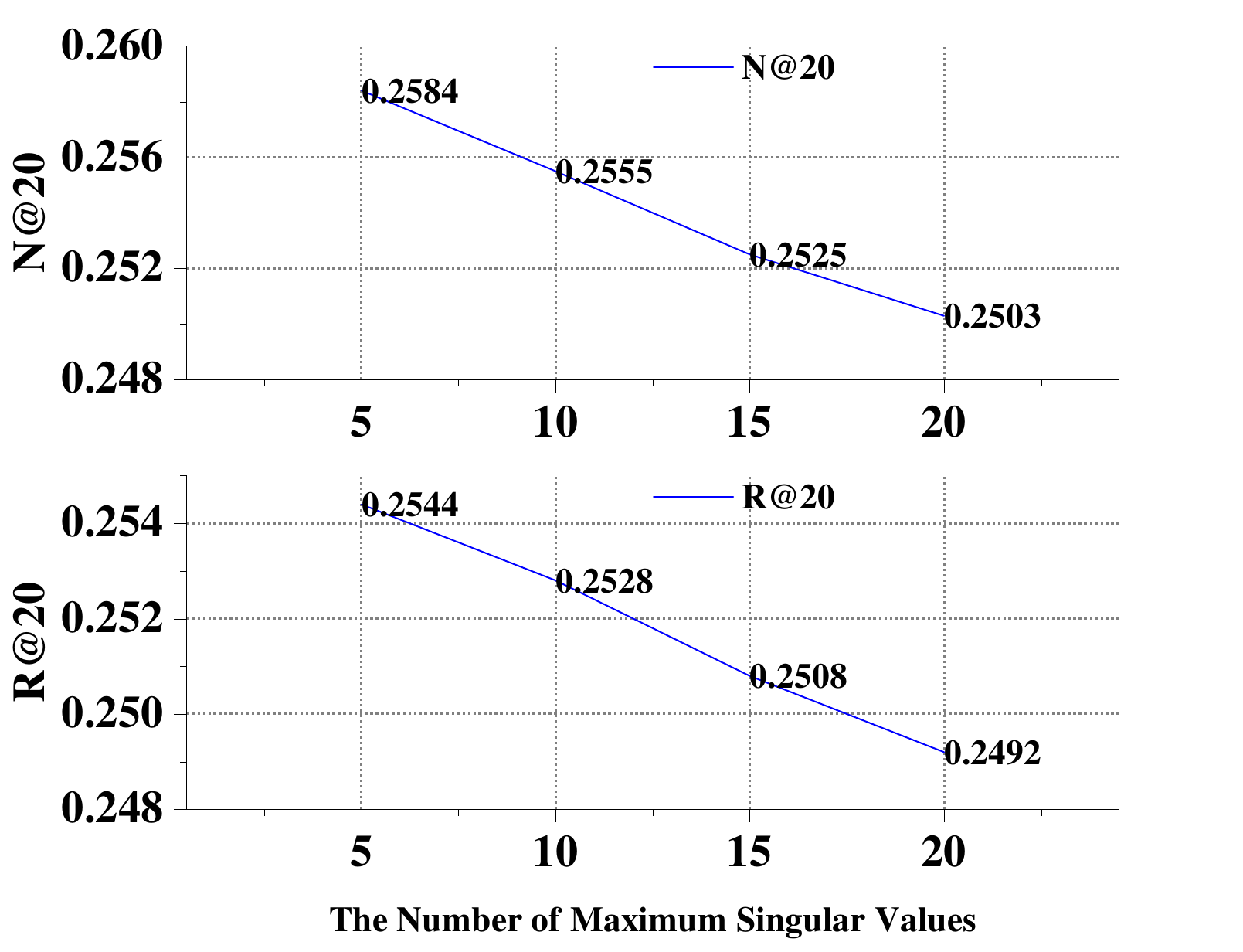}
    \caption{Impact of the number of maximum singular values(LastFM-1000) with R@20 and N@20}
    \label{Fig.7}
\end{figure}

In this section, we explore the impact of the number of singular values on the performance of the method using SVD. We conducted experiments on the ML-100K dataset and LastFM-1000 dataset, observing the changes in method performance as we adjusted the number of singular values. As illustrated in Fig.\ref{Fig.4}, Fig.\ref{Fig.5}, Fig.\ref{Fig.6} and Fig.\ref{Fig.7}, the trend of method performance with varying singular values is clearly depicted. The best performance of the method is achieved when the number of singular values is set to 5.

As shown in Fig.\ref{Fig.4} and Fig.\ref{Fig.5}, on the ML-100K dataset, the model performance curve exhibits a declining trend as the number of maximum singular values increases. The method achieves its peak performance when the number of maximum singular values is set to 5, with R@10 and N@10 being 0.2232 and 0.4034, respectively, and R@20 and N@20 being 0.3351 and 0.4008, respectively. This may indicate that increasing the number of maximum singular values does not necessarily enhance the effective information obtained by the method; rather, it can introduce noise interference, leading to a decline in performance.

The curves displayed in Fig.\ref{Fig.6} and Fig.\ref{Fig.7} illustrate a similar trend. As the number of maximum singular values increases to 20, the value of R@10 decreases from 0.1815 to 0.1714, while N@10 declines from 0.2184 to 0.2098. R@20 and N@20 also show a clear decreasing trend, with R@20 dropping from 0.2544 to 0.2492 and N@20 decreasing from 0.2584 to 0.2503. This further corroborates the significant impact of the selection of the number of maximum singular values on the performance of the method.

Experiments conducted on both datasets indicate that the number of maximum singular values in SVD is crucial for our method. Selecting an appropriate number of maximum singular values significantly impacts model performance, as an excessive number of singular values may introduce irrelevant information, thereby degrading the method's performance.

\subsubsection{Impact of number of aggregation layers}
\begin{figure}[ht]
    \centering
    \includegraphics[width=0.8\linewidth]{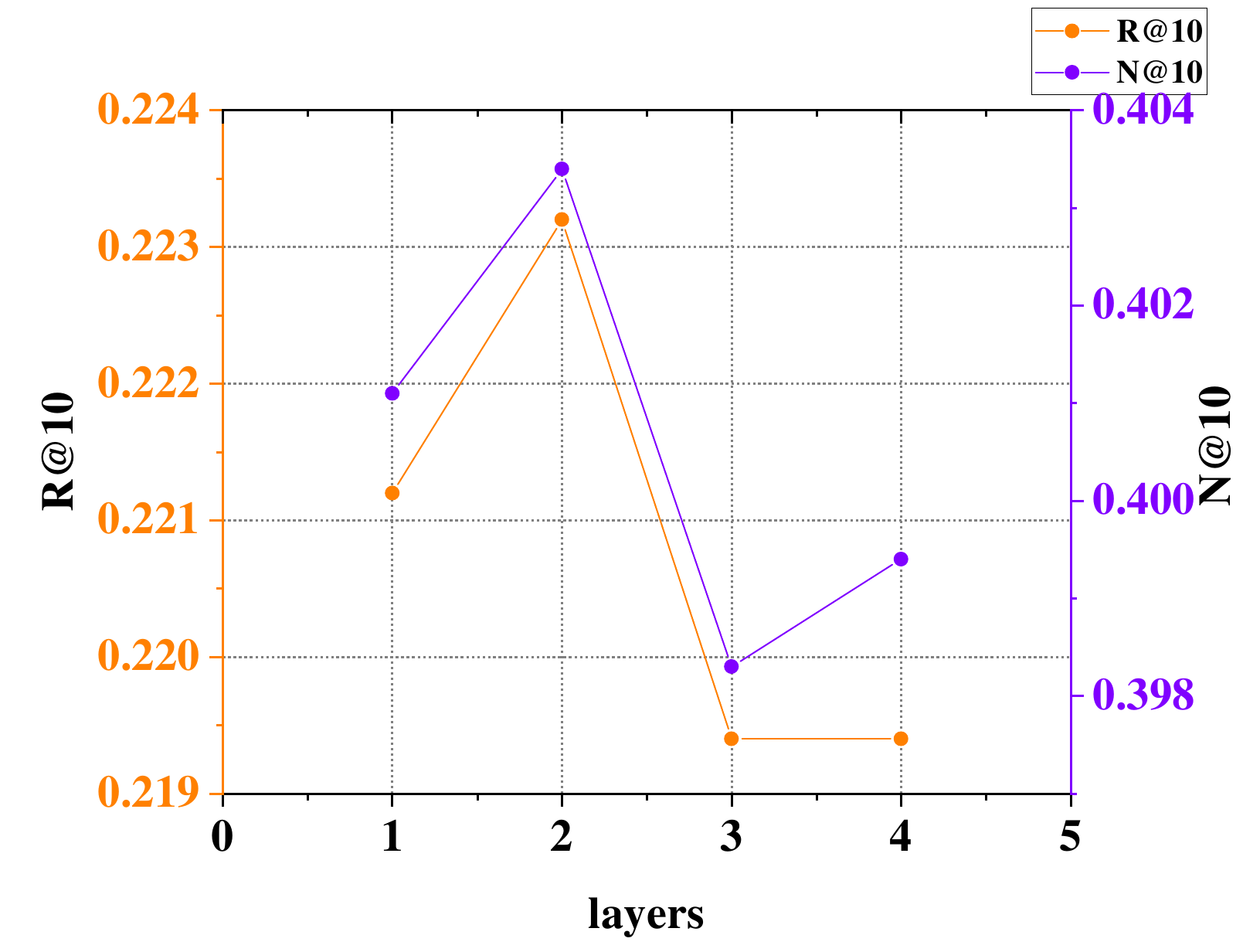}
    \caption{Impact Impact of number of aggregation layers(ML-100K) with R@10 and N@10}
    \label{Fig.8}
\end{figure}

\begin{figure}[ht]
    \centering
    \includegraphics[width=0.8\linewidth]{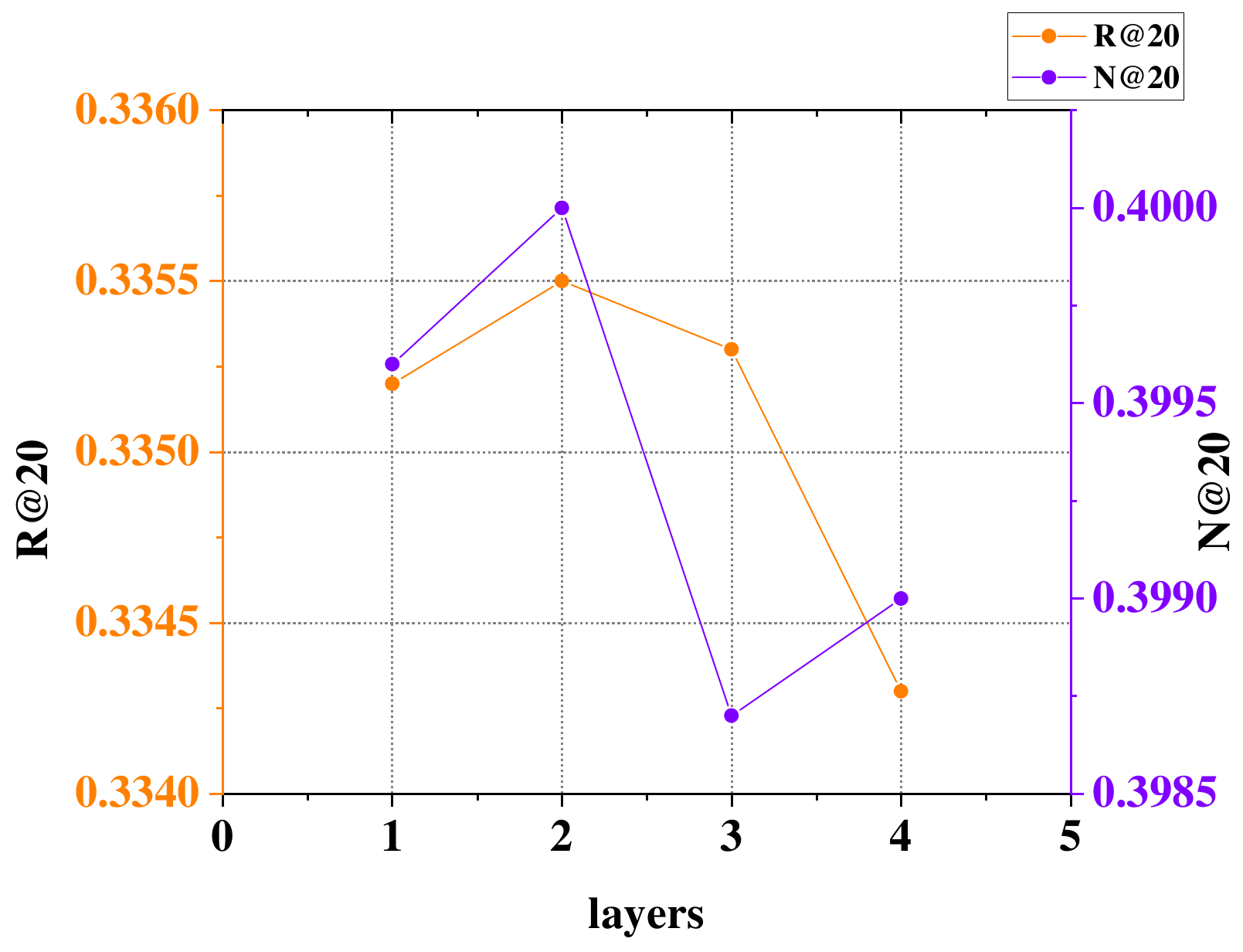}
    \caption{Impact Impact of number of aggregation layers(ML-100K) with R@20 and N@20}
    \label{Fig.9}
\end{figure}

\begin{figure}[ht]
    \centering
    \includegraphics[width=0.8\linewidth]{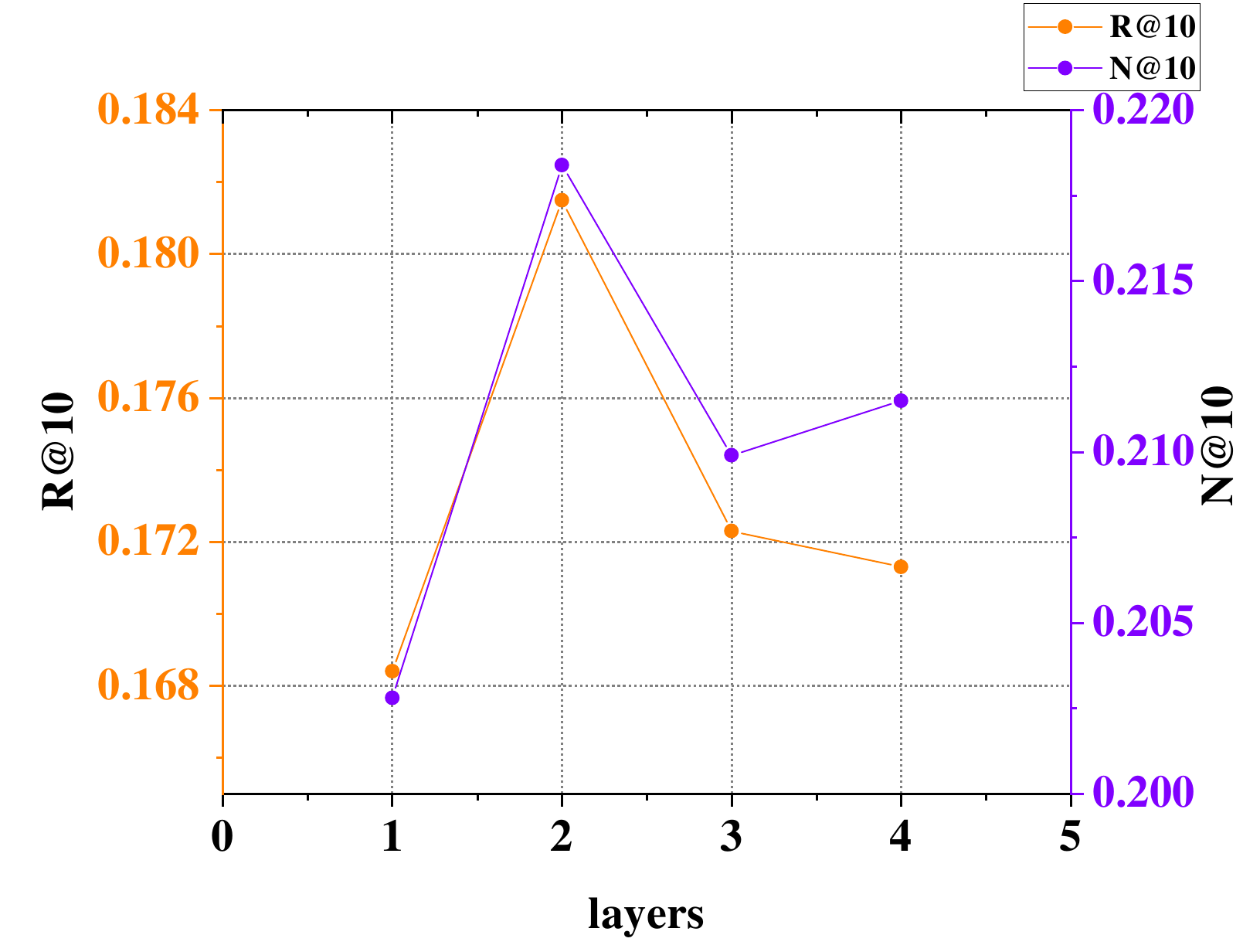}
    \caption{Impact Impact of number of aggregation layers(LastFM-1000) with R@10 and N@10}
    \label{Fig.10}
\end{figure}

\begin{figure}[ht]
    \centering
    \includegraphics[width=0.8\linewidth]{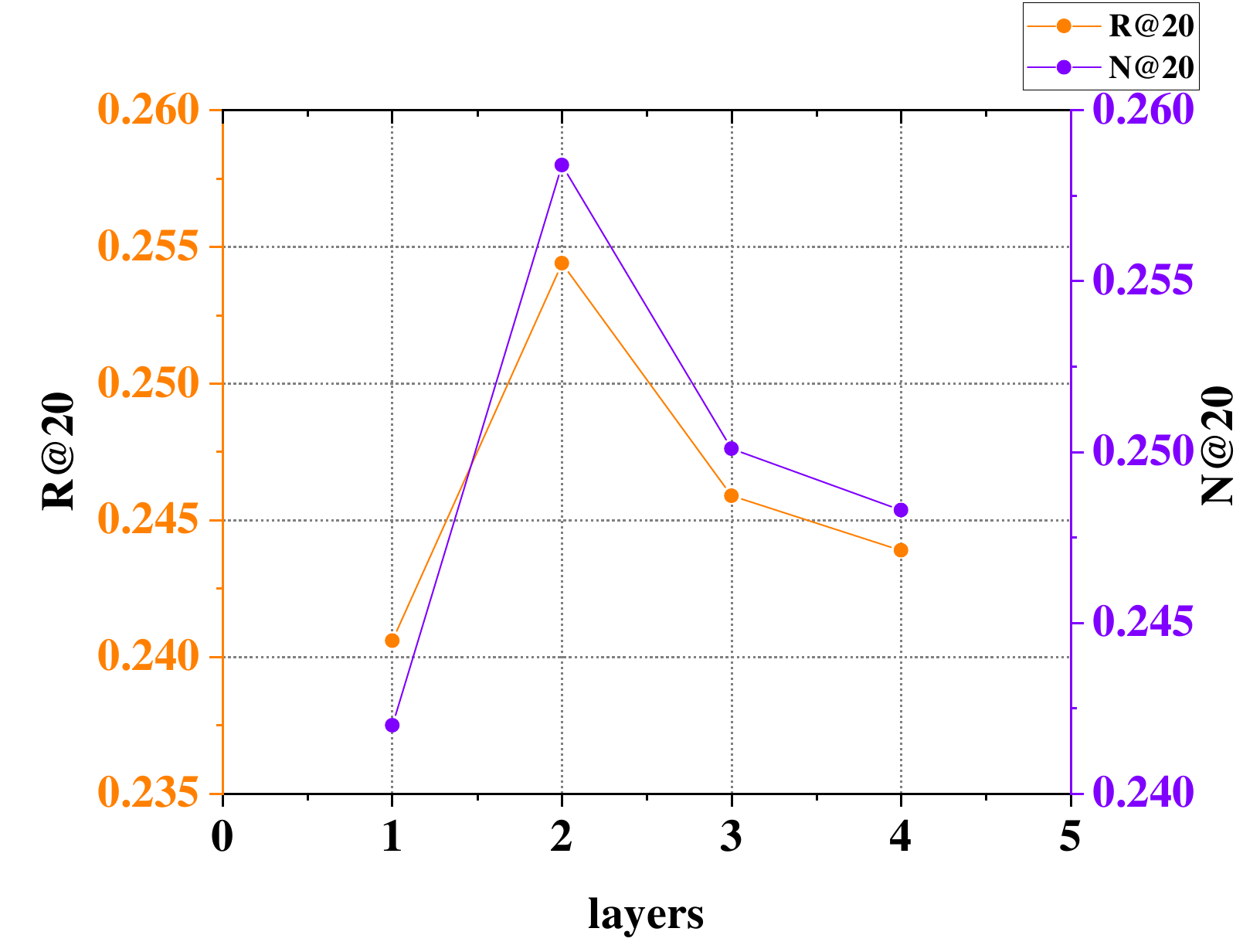}
    \caption{Impact Impact of number of aggregation layers(LastFM-1000) with R@20 and N@20}
    \label{Fig.11}
\end{figure}

We also investigated the impact of varying aggregation layer depths on the method's performance. By adjusting the number of aggregation layers, we observed changes in the method's performance on the ML-100K dataset and the LastFM dataset, as depicted in Fig.\ref{Fig.8}, Fig.\ref{Fig.9}, Fig.\ref{Fig.10} and Fig.\ref{Fig.11} respectively.

As shown in Fig.\ref{Fig.8} and Fig.\ref{Fig.9}, there are notable variations in the performance of the method on the ML-100K dataset when the number of aggregation layers is adjusted. The maximum values of R@10 and N@10 are achieved when the number of aggregation layers is set to 2, yielding values of 0.2232 and 0.4034, respectively. However, with only 1 aggregation layer, the highest values for R@20 and N@20 are 0.3352 and 0.3996, respectively. Furthermore, as the number of aggregation layers increases, the deterioration in performance becomes increasingly evident. For instance, when the number of aggregation layers is set to 4, R@10 and R@20 decrease to 0.2194 and 0.3343, respectively, while N@10 and N@20 decrease to 0.3994 and 0.399.

As illustrated in Fig.\ref{Fig.10} and Fig.\ref{Fig.11}, on the LastFM-1000 dataset, the variations in the method's performance with changes in the number of aggregation layers exhibit a similar overall trend to that observed on the ML-100K dataset. The maximum values of R@10 and N@10 are achieved when the number of aggregation layers is set to 2, yielding values of 0.1815 and 0.2184, respectively. However, with only 1 aggregation layer, the values for R@10 and N@10 are 0.1684 and 0.2028, respectively. Additionally, as the number of aggregation layers increases, performance also deteriorates. For example, when the number of aggregation layers is set to 4, R@10 and R@20 decrease to 0.1713 and 0.2439, respectively, while N@10 and N@20 decrease to 0.2115 and 0.2483. This might suggest that, as the number of recommended items increases, adding more aggregation layers could introduce excessive noise or irrelevant information, leading to a decline in performance. This indicates that an excessive number of aggregation layers might capture redundant information, resulting in an overly smooth information acquisition process, which ultimately leads to a decline in method performance.

The examination of the experimental results indicates that the depth of the aggregation layers in graph neural networks has a relatively significant impact on recommendation performance. Both excessively small and excessively large aggregation layer depths typically have negative effects on performance. Selecting an appropriate aggregation layer depth is crucial for improving the method's performance.

\subsubsection{Impact of embedding size}
\begin{figure}[ht]
    \centering
    \includegraphics[width=0.8\linewidth]{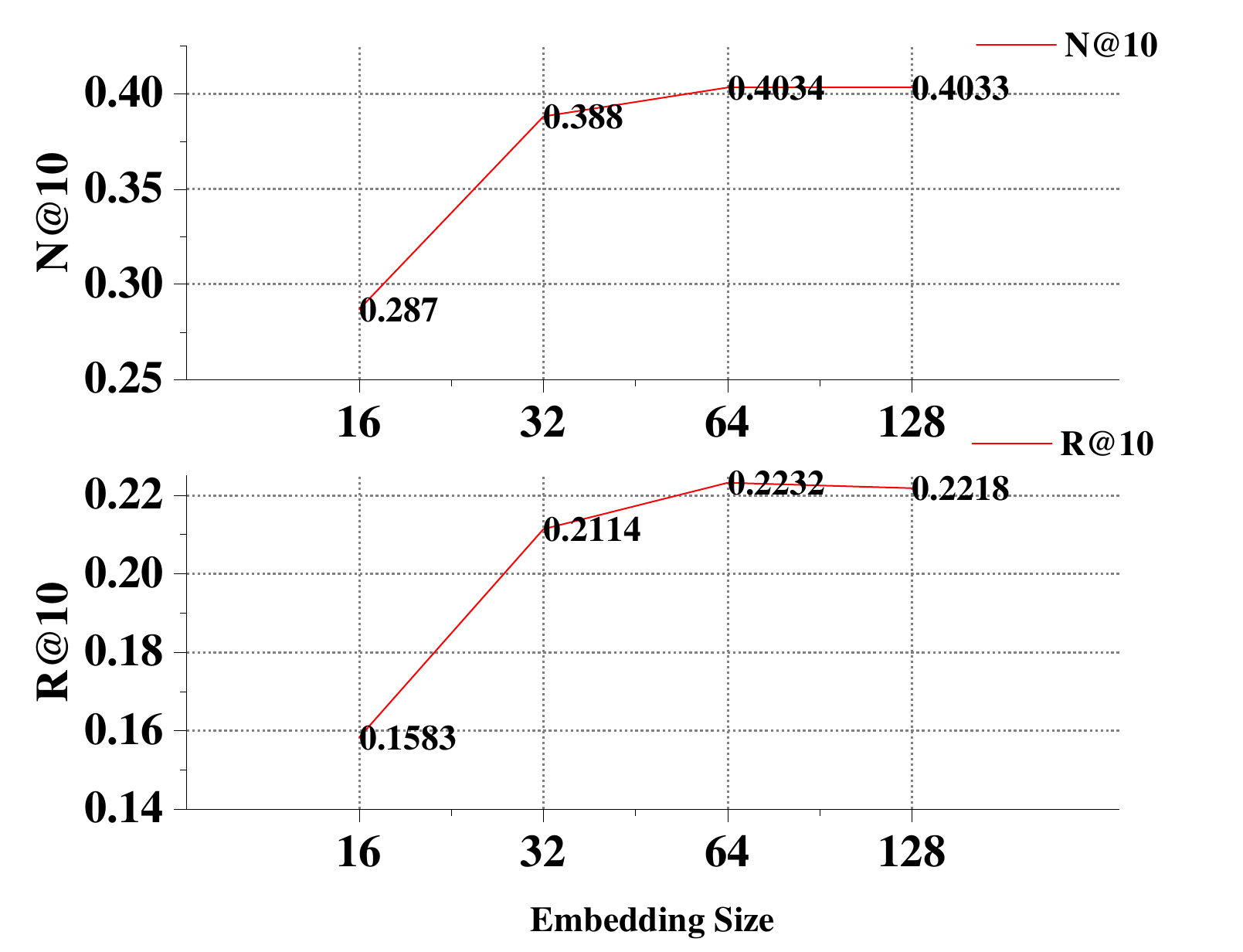}
    \caption{Impact of embedding size(ML-100K) with R@10 and N@10}
    \label{Fig.12}
\end{figure}

\begin{figure}[ht]
    \centering
    \includegraphics[width=0.8\linewidth]{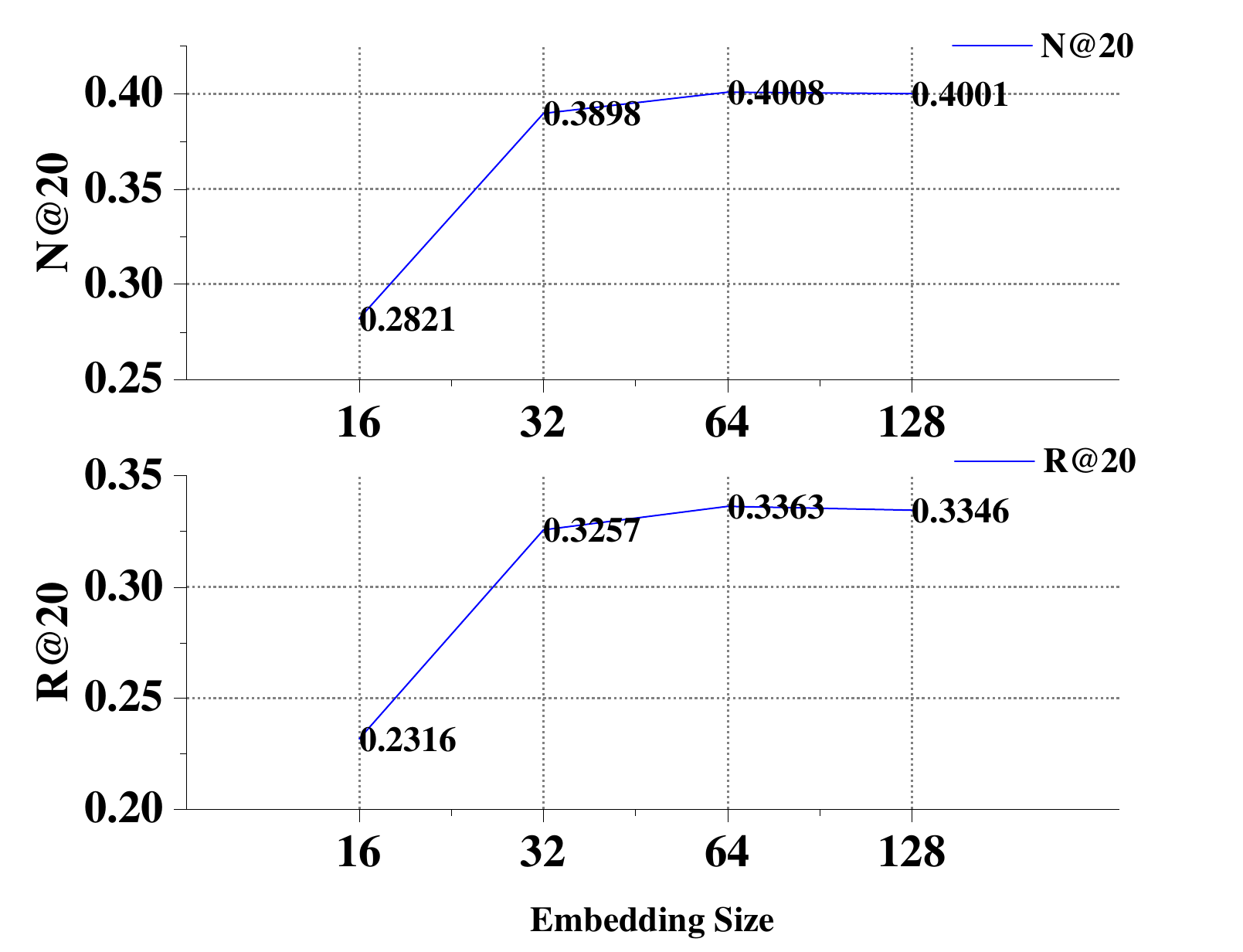}
    \caption{Impact of embedding size(ML-100K) with R@20 and N@20}
    \label{Fig.13}
\end{figure}

\begin{figure}[ht]
    \centering
    \includegraphics[width=0.8\linewidth]{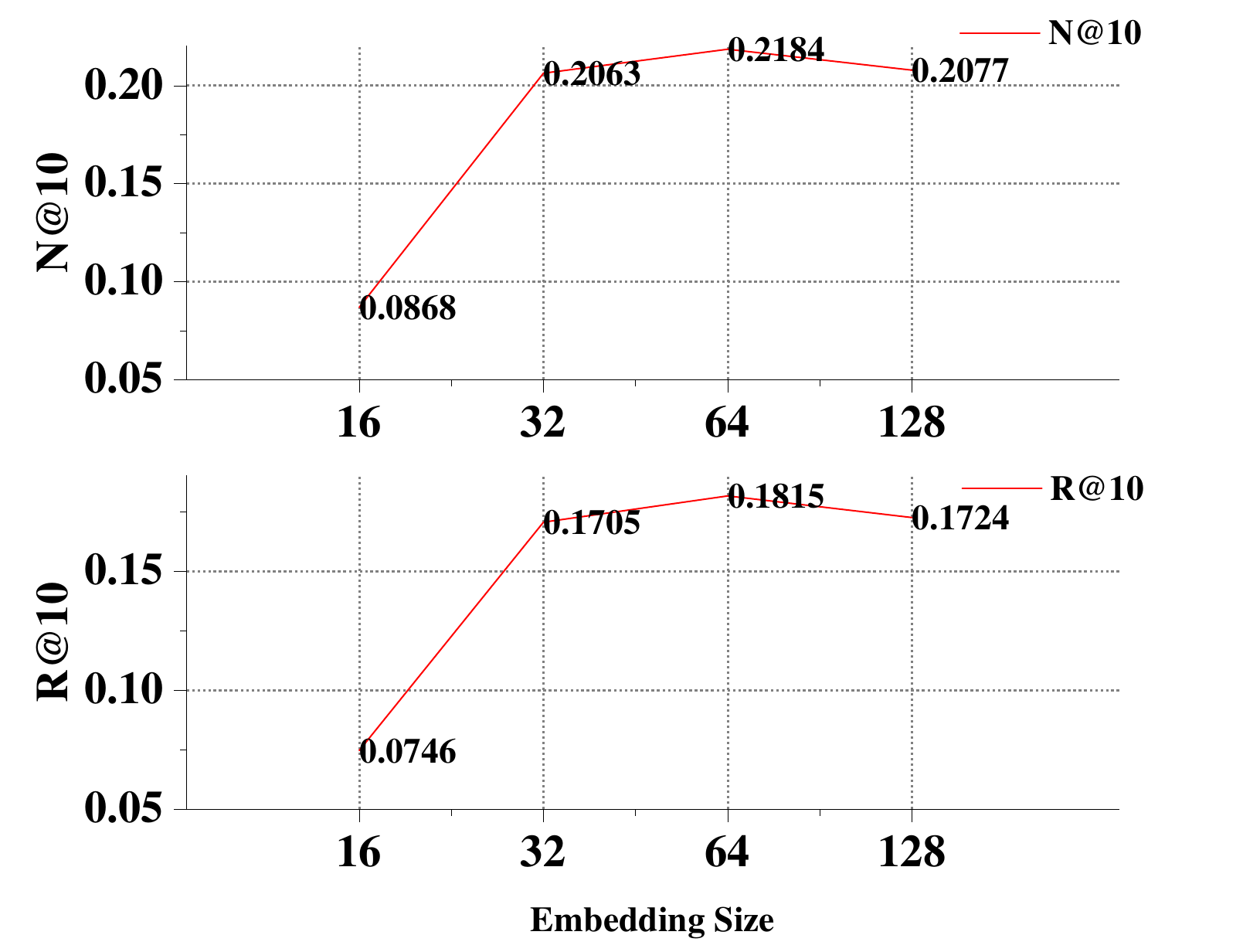}
    \caption{Impact of embedding size(LastFM-1000) with R@10 and N@10}
    \label{Fig.14}
\end{figure}

\begin{figure}[ht]
    \centering
    \includegraphics[width=0.8\linewidth]{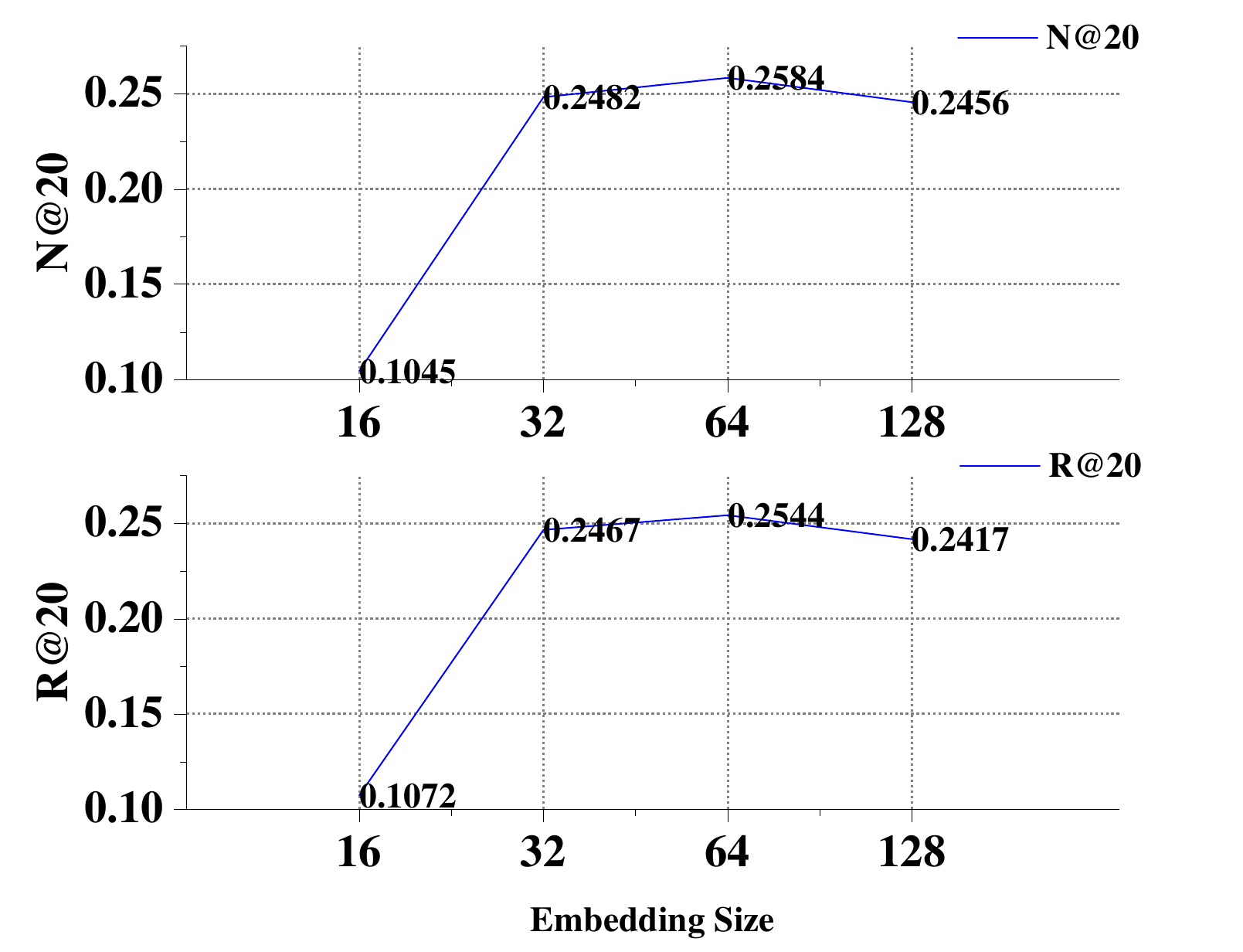}
    \caption{Impact of embedding size(LastFM-1000) with R@20 and N@20}
    \label{Fig.15}
\end{figure}

In this section, we fine-tuned the embedding dimension in the method on the ML-100K and LastFM-1000 datasets to observe the trend of method performance changes. As indicated by the evaluation metric curves in Fig.\ref{Fig.12}, Fig.\ref{Fig.13}, Fig.\ref{Fig.14} and Fig.\ref{Fig.15}, the method achieves its best performance when the embedding dimension is 64. Performance deteriorates to varying degrees when the embedding dimension either increases or decreases.

On the ML-100K dataset, as shown in Fig.\ref{Fig.12} and Fig.\ref{Fig.13}, the method generally exhibits an upward trend. When the embedding dimension increases to 64, the method reaches its maximum performance, with R@10 and N@10 achieving 0.2232 and 0.4034, respectively, and R@20 and N@20 achieving 0.3363 and 0.4008, respectively. Continuing to increase the dimension to 128 results in a slight decline in performance, with R@10 and N@10 dropping to 0.2218 and 0.4033, respectively, and R@20 and N@20 dropping to 0.3346 and 0.4001, respectively. The minimum performance is observed at an embedding dimension of 16, where the performance deteriorates significantly, with R@10 and N@10 only reaching 0.1583 and 0.287, respectively, and R@20 and N@20 reaching 0.2316 and 0.2821, respectively.

As shown in Fig.\ref{Fig.14} and Fig.\ref{Fig.15}, the evaluation metric curves on the LastFM-1000 dataset show a similar trend. As the embedding dimension increases from 16 to 64, the performance gradually improves to its peak, with R@10 increasing from 0.074 to 0.1851 and N@10 increasing from 0.0868 to 0.2184. When the embedding dimension increases to 128, a slight decrease in performance is observed, with R@10 and N@10 decreasing to 0.1724 and 0.2077, respectively, and R@20 and N@20 decreasing to 0.2417 and 0.2456, respectively. The changes in the curves suggest that an appropriate dimension helps the method capture accurate information. Dimensions that are too small lack the ability to gather sufficient information, leading to a loss of useful information and a resultant decline in method performance. Conversely, excessively large dimensions compensate for the lack of information but also face issues of redundancy. Unnecessary information can interfere with the method's ability to recognize relevant information, causing the method to accept extraneous data and thereby impairing its performance.

The experimental results above demonstrate that the embedding dimension has a significant impact on the performance of recommendation methods based on graph neural networks. Both excessively large and excessively small embedding dimensions tend to compromise the performance of the recommendation methods. Therefore, selecting an appropriate embedding dimension has a positive effect on the improvement of method performance.

\subsection{Loss analysis}
\begin{figure}[ht]
    \centering
    \includegraphics[width=0.8\linewidth]{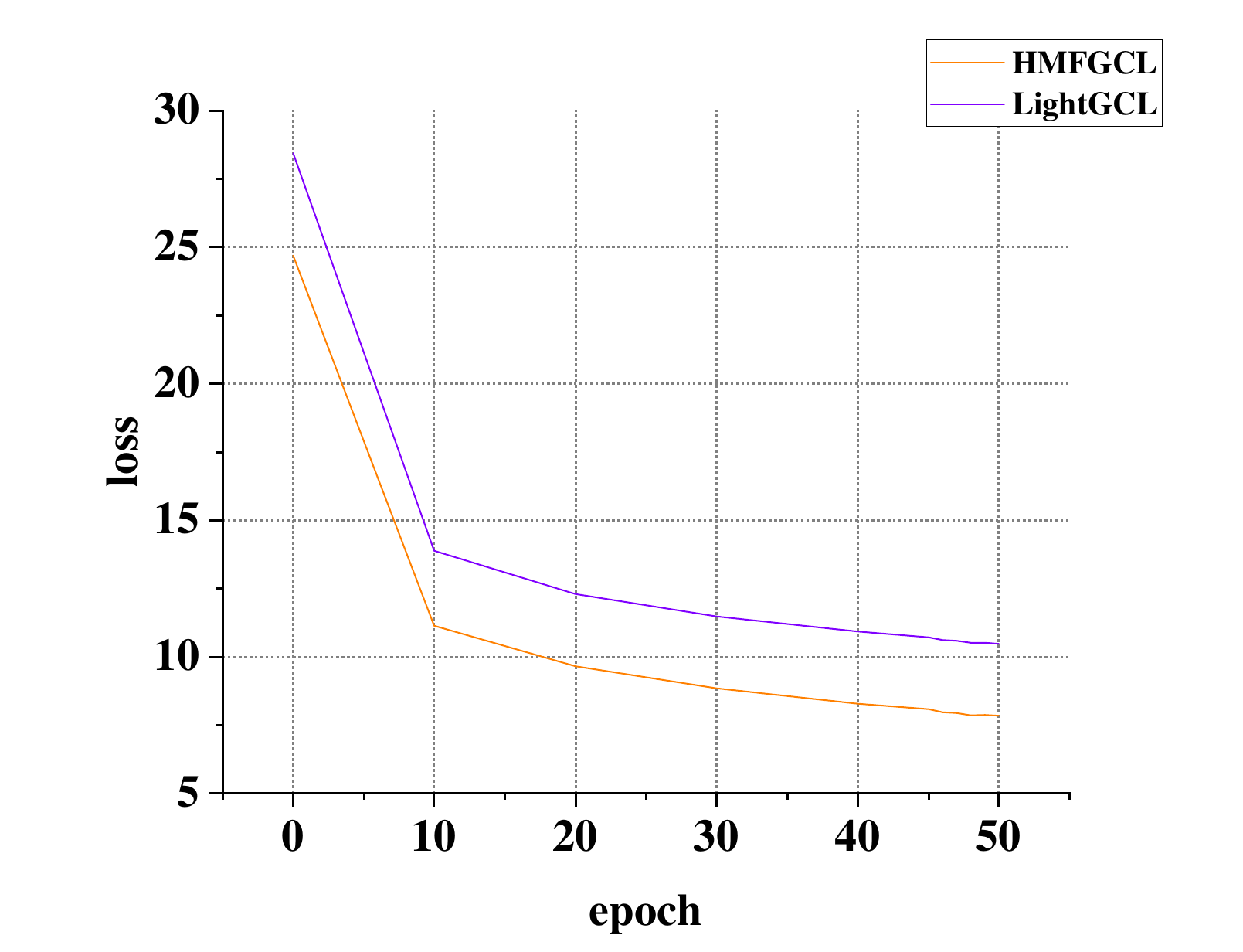}
    \caption{Loss Analysis(ML-100K)}
    \label{Fig.16}
\end{figure}

\begin{figure}[ht]
    \centering
    \includegraphics[width=0.8\linewidth]{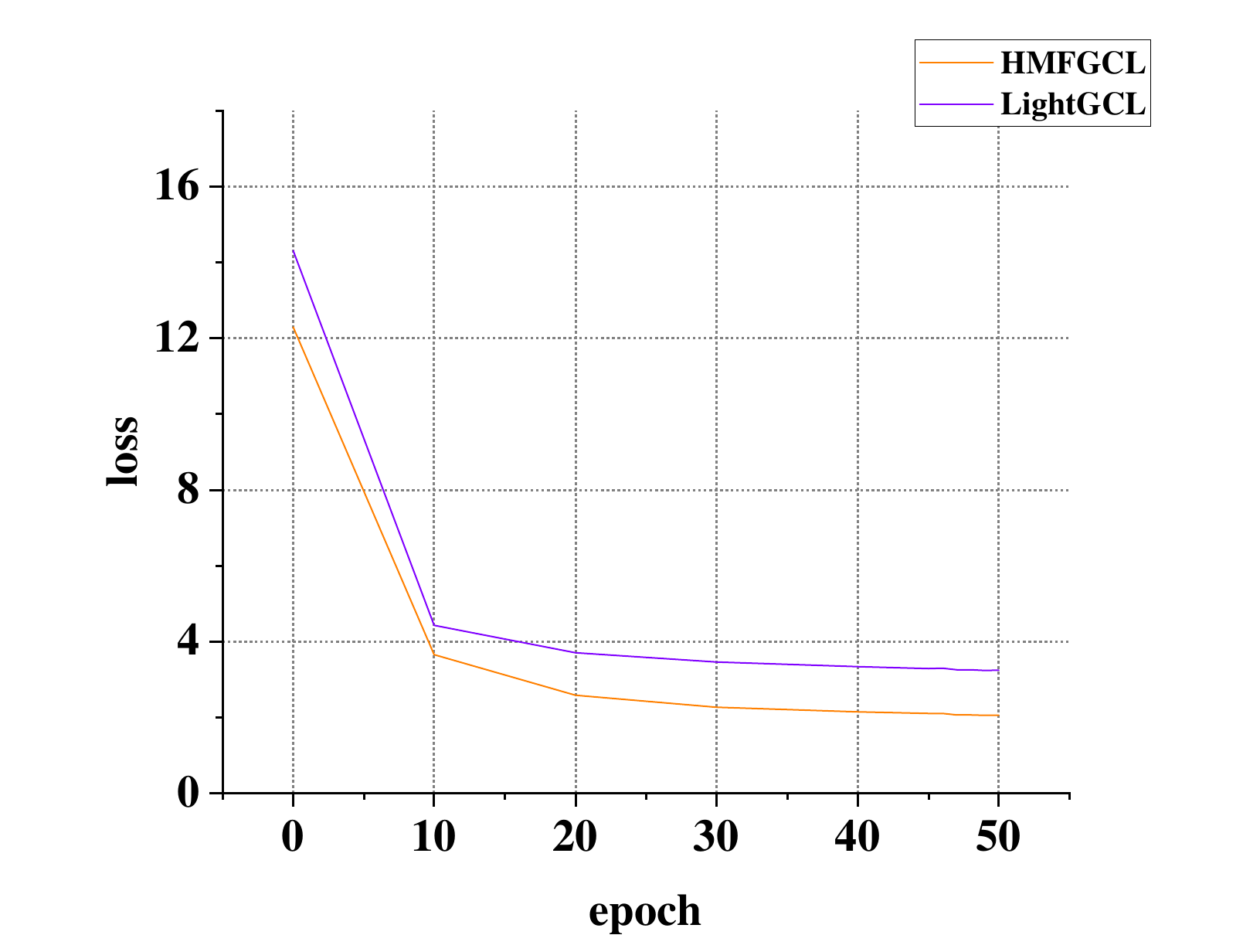}
    \caption{Loss Analysis(LastFM-1000)}
    \label{Fig.17}
\end{figure}

In this section, we investigated the changes in loss values in the ML-100K and LastMF-1000 datasets, and conducted a comparison and analysis, as illustrated in Fig.\ref{Fig.16} and Fig.\ref{Fig.17}. We selected the first 50 epochs as samples for analysis while choosing the suboptimal baseline, LightGCL, as a comparative benchmark to observe the convergence trend of the loss values.

As shown in Fig.\ref{Fig.16}, our method demonstrates superior performance on the ML-100K dataset, with a slightly faster-declining trend and lower loss values compared to the suboptimal baseline. This indicates its exceptional learning capability and improved generalization ability. In comparison to LightGCL, our approach also exhibits a favorable declining trend, although its loss values are higher than those of our method.

As depicted in Fig.\ref{Fig.17}, on the ModCloth-1000 dataset, our method also achieves lower loss values compared to the suboptimal baseline. Both our method and LightGCL exhibit equally favorable declining trends, with the decline being quite apparent. This might suggest that both methods demonstrate good generalization capabilities on this dataset and effectively avoid overfitting.

The analysis of the experimental results indicates that the design of the loss function plays a crucial role in graph-based contrastive learning recommendation methods. A faster-declining trend and lower loss values are essential for enhancing the performance of the recommendation method. How to construct the model architecture and design a more reasonable loss function is of significant importance for improving the performance of the method.

\subsection{Ablation experiment}
Here, we investigate the impact of the hybrid matrix factorization module on the performance of the method on the ML-100K and LastFM-1000 datasets. As shown in Table \ref{Tab.3}, we present the performance differences among three variants: HMFGCL-M (which retains only the MF module), HMFGCL-S (which retains only the SVD module), and HMFGCL-R (which retains neither module).

\begin{table}[htbp]
\centering
\caption{We conducted experiments on the two datasets using the original method and the three variants, employing Recall and NDCG as evaluation metrics with k set to 10 and 20, denoted as R@10, R@20, N@10, and N@20.}
\label{Tab.3}
\resizebox{0.5\textwidth}{!}{
\begin{tabular}{l|cccc|cccc}
\hline
& \multicolumn{4}{c|}{\textbf{ML-100K}} & \multicolumn{4}{c}{\textbf{LastFM-1000}} \\
\textbf{Method} & \textbf{R@10}  & \textbf{R@20}  & \textbf{N@10}  & \textbf{N@20}  & \textbf{R@10}  & \textbf{R@20}  & \textbf{N@10}  & \textbf{N@20} \\ \hline
\textbf{HMFGCL-R} & 0.166 & 0.2505 & 0.3051 & 0.301 & 0.1669 & 0.2357 & 0.2026 & 0.2406 \\
\textbf{HMFGCL-M} & 0.2168 & 0.3287 & 0.395 & 0.3928 & 0.1646 & 0.2321 & 0.1979 & 0.2351 \\
\textbf{HMFGCL-S} & 0.2133 & 0.3283 & 0.3858 & 0.3862 & 0.1728 & 0.2486 & 0.2086 & 0.25 \\
\textbf{HMFGCL} & 0.2232 & 0.334 & 0.4034 & 0.4 & 0.1815 & 0.2544 & 0.2184 & 0.2584 \\ \hline
\end{tabular}
}
\end{table}

As shown in Table \ref{Tab.3}, overall, both HMFGCL-M (retaining only the MF) and HMFGCL-S (retaining only the SVD) show improvements, and retaining both achieves the maximum performance, indicating the effectiveness of the module on both the ML-100K and LastFM-1000 datasets. The experimental results on the ML-100K dataset show that HMFGCL-R has the lowest performance, with R@10 and R@20 being only 0.166 and 0.2505, respectively, and N@10 and N@20 being only 0.3051 and 0.301, respectively, indicating a significant performance decrease. In contrast, both HMFGCL-M and HMFGCL-S show noticeable improvements, with R@10 increasing to 0.2168 and 0.2133, respectively, and N@10 increasing to 0.395 and 0.3858, respectively.

The results on the LastFM-1000 dataset indicate that the removal of the module leads to a significant decline in the metrics for HMFGCL-R, with R@10 dropping from 0.1815 to 0.1669 and N@10 decreasing from 0.2184 to 0.2026. Furthermore, the performance of HMFGCL-M and HMFGCL-S also shows notable declines, with R@10 decreasing to 0.1646 and 0.1728, respectively, and N@10 dropping to 0.1979 and 0.2086, respectively. The experimental data demonstrate that the hybrid matrix factorization module we employed is effective for both public datasets and serves to positively assist in the integration of global collaborative information in the recommendation methods, thereby enhancing the performance of the approach.

\section{Conclusion}

In this paper, we propose a graph-based contrastive learning recommendation method that integrates global information through a fusion of Matrix Factorization (MF) and Singular Value Decomposition (SVD) techniques, combined with mixed noise perturbation. This method first utilizes both MF and SVD to capture global information between user-item pairs and then fuses these methods to generate enhanced views for contrastive learning. Additionally, we enhance the uniformity and quality of embedding representations by incorporating mixed noise during the forward propagation of graph neural networks. Extensive experiments conducted on three public datasets demonstrate the superiority of our approach compared to baseline methods in specific scenarios, particularly in terms of effectiveness on small datasets. Future research holds several intriguing directions. Firstly, the issue of dataset size should be addressed to improve efficiency and performance when dealing with large datasets. Secondly, exploring new methods for generating contrasting views or fusing multiple enhanced views for contrastive learning is an interesting avenue. Lastly, while graph neural networks can provide speculative information for cold-start users or items, finding better solutions to mitigate the cold-start problem remains an important research area.

\end{document}